\documentclass[12pt]{article}
\usepackage{epsfig}
\usepackage{amsfonts}
\usepackage{amssymb}
\usepackage{amsmath}
\usepackage{amsthm}
\usepackage{latexsym}
\usepackage{graphicx}
\usepackage{bm}
\usepackage{tabularx}
\usepackage{booktabs} 
\usepackage{enumerate}
\usepackage{caption}
\usepackage{wrapfig}  

\usepackage[utf8]{inputenc}
\usepackage[T1]{fontenc}

\usepackage[numbers]{natbib}
\usepackage[titletoc]{appendix}
\usepackage{comment}

\usepackage{bbm} 
\usepackage{url}
\usepackage{subfig}
\usepackage{hyperref}
\usepackage{color}
\usepackage{float}
\catcode`\@=11
\newcommand{\shortdot}[1]{\raisebox{-0.4pt}{$\stackrel{\bullet}{#1}$}}
\theoremstyle{plain}
\newtheorem{theorem}{Theorem}[section]

\newtheorem{corollary}[theorem]{Corollary}

\theoremstyle{definition}

\theoremstyle{remark}
\newtheorem*{remark}{Remark}

\title{Flattening the Curve: Insights From Queueing Theory}
\author{ 
  Sergio Palomo \\ Systems Engineering \\ Cornell University
\\ 418 Upson Hall, Ithaca, NY 14853 \\  sdp85@cornell.edu \\ \and
  Jamol Pender \footnote{Corresponding Author} \\ School of Operations Research and Information Engineering \\ Cornell University
\\ 228 Rhodes Hall, Ithaca, NY 14853 \\  jjp274@cornell.edu \\ \and
  William Massey \\ Department of Operations Research and Financial Engineering \\Princeton University
\\ 206 Sherrerd Hall, Princeton, NJ 08540 \\  wmassey@princeton.edu  \\ 
\and 
Robert C. Hampshire \\ Gerald R. Ford School of Public Policy \\ University of Michigan
\\ 735 S. State Street, Ann Arbor, MI 48019 \\  hamp@umich.edu
 }
\begin{document}

\maketitle

\begin{abstract}
The worldwide outbreak of the coronavirus  was first identified in 2019 in Wuhan, China. Since then, the disease has spread worldwide. As it currently spreading in the United States, policy makers, public health officials and citizens are racing to understand the impact of this virus on the United States healthcare system. They fear that the rapid influx of patients will overwhelm the healthcare system leading to unnecessary fatalities. Most countries and states in America have introduced mitigation strategies, such as social distancing, to decrease the rate of newly infected people, i.e. flattening the curve. 

In this paper, we analyze the time evolution of the number of people hospitalized due to the coronavirus using the methods of queueing theory.  Given that the rate of new infections varies over time as the pandemic evolves, we model the number of coronavirus patients as a  dynamical system  based on the theory of infinite server queues with non-stationary Poisson arrival rates. With this model we are able to quantify how flattening the curve affects the peak demand for hospital resources. This allows us to characterize how aggressively society must flatten the curve in order to avoid overwhelming the capacity of healthcare system.  We also demonstrate how flattening the curve impacts the elapsed time between the peak rate of hospitalizations and the time of the peak demand for the hospital resources. Finally, we present empirical evidence from China, South Korea, Italy and the United States that supports the insights from the model.  
        \\

%\textbf{Corresponding Author:  Jamol Pender} \\

%AMS subject classifications: 34K40, 34K18, 41A10, 37G15, 34K27 %, 35Q94, 41A10, 37G15 %34K99, 35Q94, 41A10, 37G15

\end{abstract}

%**************************************************************************
%**************************************************************************

\section{Introduction}

The coronavirus was first identified in Wuhan, a city of over 11 million people in China's Hubei province, in December of 2019. COVID-19, the disease caused by the coronavirus, grew by several thousand per day in China between late January and the peak of their epidemic in early February. The number of infections appearing each day has decreased there significantly. This was mainly due to stringent containment efforts. 

However, the outbreak is now a global pandemic. Large outbreaks in South Korea, Iran, Italy, Spain, and the United States have all sparked interest in coronavirus research. As of April 19th 2020, there have been 2,432,092 cases and 166,794 deaths confirmed worldwide. Also, the World Health Organization has officially named COVID-19 a global pandemic. The spread of the virus has been rapid, with 208 countries now having reported cases. 

Lawmakers and public health experts such as National Institute of Allergy and Infectious Diseases director, Dr. Anthony Fauci, have been encouraging “social distancing” as a measure to flatten the curve. This implies that we want to push down the peak number of infected people and spread it out over time. Many people understand how flattening the curve impacts the healthcare system from an empirical or qualitative perspective, see \citet{goebel2020}. However, there is little quantitative research in this area and relatively little research on how much pushing is needed to flatten the curve. 

In this paper, we model the potential demand for hospital resources like beds, ventilators, medical staff as an infinite server queue with non-stationary arrival rates. We use as our simple examples of unimodal arrival rate functions, the scaled Gaussian distribution density and the scaled gamma density functions.  We then derive explicit expressions for the peak mean number in the queueing system as well as the time of this peak.

Our paper has two main goals. The first one is showing the global community how operations research, represented here by queueing theory, helps us understand the impact of COVID-19 on the healthcare system. The other goal is showing that by leveraging insights from queues with time varying rates leads to simple descriptions of the hospital patient dynamics for the COVID-19 pandemic. More specifically, we show using a scaled Gaussian arrival rate and a scaled Gamma arrival rate, how to calculate the time of the peak infection rate, the peak mean number of infected people, and the time between the peak hospitalization rate and the time of the peak number of patients. We use an infinite server queueing model to demonstrate the impact of social distancing and flattening the curves of arrival rates and mean counts of on hospital capacity. 

Operationally, we can learn from the situations we have already observed in China, Europe, and New York City to avoid similar outcomes in other places? From the data, we see that "every hotspot has its own curve" see \citet{Hilk}.  One important advantage of our work is that it can be used to study each individual curve for individual regions around the world. The geographic and temporal clustering of outbreaks have the potential to create a health care system collapse. Many states in the United States have imposed "stay at home" orders to contain the spread of the disease. However, since many people are asymptomatic, we can show through the analysis of our queueing examples that containment measures implemented in recent days may take weeks to have an observable effect in the data. 

Although hospital bed capacity is an important concern for health care officials, it is not the most crucial bottleneck for providing critical care when patients need it. After observing many deaths in Italy, many experts and government officials are concerned about the number of available ventilators. Patients who become critically ill need individual ventilators but they have become a scarce resource. Equally important resources are medical specialists and staff members that can supervise the patients in their beds and operate the ventilators properly. Fundamentally, this paper strives to help healthcare managers understand what supply is needed for all these lifesaving resources at the peak of the ultimate curve for their aggregate demand. 

\subsection{Main Contributions of This Paper}

The contributions of this work can be summarized as follows:    
\begin{itemize}
\item Given a non-stationary arrival rate, we derive an expression for the mean number of infected individuals using an infinite server queue.  
\item Using the infinite server queue model, we derive the exact time of the peak load and the value of the peak load.  
\item Using our peak load calculations, we determine the value of "flattening" that needs to be done in order to make the peak load infection value below a pre-specified capacity level.  
\item We study the impact of the duration of the virus on the peak load dynamics of hospital resources and how it affects our flattening policies. 
\item We characterize the nonlinear relationship between flattening the curve and the time lag between the peak rate of newly admitted patients and the peak demand for hospital resources.  We also show how the lag can be observed in real data.  
\end{itemize}

%**************************************************************************
%**************************************************************************
%
%\subsection{Organization of Paper}
%
%The remainder of this paper is organized as follows.   

\section{Demand for Hospital Resources: The Infinite Server Queue}

We model the number of patients as a $M_t/G/\infty$ queueing model.  There are two important reasons to begin with the $M_t/G/\infty$ queueing model even though it has an infinite number of servers and this does not seem realistic from a practical perspective.  The first reason is that the $M_t/G/\infty$ queue is very tractable since the distribution is known in closed form for any non-stationary arrival rate.  The second reason is that the $M_t/G/\infty$ queue is the best type of queue one can hope for where everyone is immediately served and no one ever waits for service.  From the perspective of the COVID-19 epidemic, this means that anyone who wants a bed gets a bed immediately or who ever needs a ventilator gets a ventilator immediately.  In this regard, the $M_t/G/\infty$ infinite server queue is an upper bound for queueing models with a finite number of servers and without abandonment since it represents the dynamics if the manager had access to an infinite amount of resources and is not resource constrained.   Moreover, the infinite server queue has a history of being used to staff finite server systems, see for example \citet{JMMW, FMMW, LW, MP3}.  Thus, the infinite server queue serves as the first step to understanding more complicated systems.  

\subsection{The $M_t/G/\infty$ Queue}
In this section, we state the closed form formulas for the $M_t/G/\infty$ queueing model, which exploits the results of \citet{EMW, EMW2} for the time varying infinite server queue.  In the paper of \citet{EMW}, they use the properties of the Poisson arrival process and use Poisson random measure arguments to show that the $M_t/G/\infty$ queue $Q^{\infty}(t)$, has a Poisson distribution with time varying mean $q^{\infty}(t)$.  As observed in \citet{EMW, EMW2}, $q^{\infty}(t)$ has the following integral representation
\begin{eqnarray}
q^{\infty}(t) &=& E[Q^{\infty}(t) ] \\
&=& \int^{t}_{-\infty} \overline{G}(t-u) \lambda(u) du \\
&=& E\left[\int^{t}_{t-S} \lambda(u) du \right] \\
&=& E[ \lambda( t - S_e ) ] \cdot E[S]
\end{eqnarray}
where $\lambda(u)$ is the time varying arrival rate and $S$ represents a service time with distribution G, $\overline{G} = 1 - G(t) = 
\mathbb{P}( S > t)$, and $S_e$ is a random variable with distribution that follows the 
stationary excess of residual-lifetime cdf $G_e$, defined by
 \begin{eqnarray}
G_e(t) &\equiv&  \mathbb{P}( S_e < t) = \frac{1}{E[S]} \int^{t}_{0} \overline{G}(u) du = \frac{1}{E[S]} \int^{t}_{0} \mathbb{P}( S > u) du, \ 
\ \ t \geq 0.
\end{eqnarray}

\begin{remark}
As a point of emphasis, we would like to remind readers that the word \emph{curve} in this paper maps to two different things: the arrival rate function and the mean offered load. The general public might call these the infection rate and the number of patients in the hospital. Moreover, there is a causal relationship between the two curves, i.e. you need to flatten the first one in order to flatten the second one.
\end{remark}

\subsection{The $M_t/G/\infty$ Queue with a Gaussian Arrival Rate}
In this section, we describe the dynamics of the  $M_t/G/\infty$ queue whose arrival rate is driven by a Gaussian distribution function.   Thus, the arrival rate function $\lambda(t)$ is given by the following expression
\begin{eqnarray}
\lambda (t) \equiv \frac{\lambda }{\sigma } \cdot \varphi \left( {\frac{{t - \tau }}{\sigma }} \right),
\end{eqnarray}  
where $\tau$ is mean of the Gaussian distribution, $\sigma$ is the standard deviation of the Gaussian distribution, $\lambda^*$ is the number of total expected number of infected individuals over the lifetime of the epidemic, and  $\varphi \left( x \right)$ is the probability density function of the standard Gaussian distribution i.e.
\begin{equation} \label{norm_bound}
\varphi \left( x \right) \equiv \frac{1}{{\sqrt {2\pi } }} \cdot {e^{ - {x^2}/2}} .
\end{equation}
One of the important things to observe is that the standard Gaussian pdf is bounded from above i.e.
\begin{equation} \label{norm_bound}
\varphi \left( x \right) \equiv \frac{1}{{\sqrt {2\pi } }} \cdot {e^{ - {x^2}/2}} \leq \frac{1}{{\sqrt {2\pi } }}.
\end{equation}

\begin{figure}[ht] 
	\includegraphics[scale =.25]{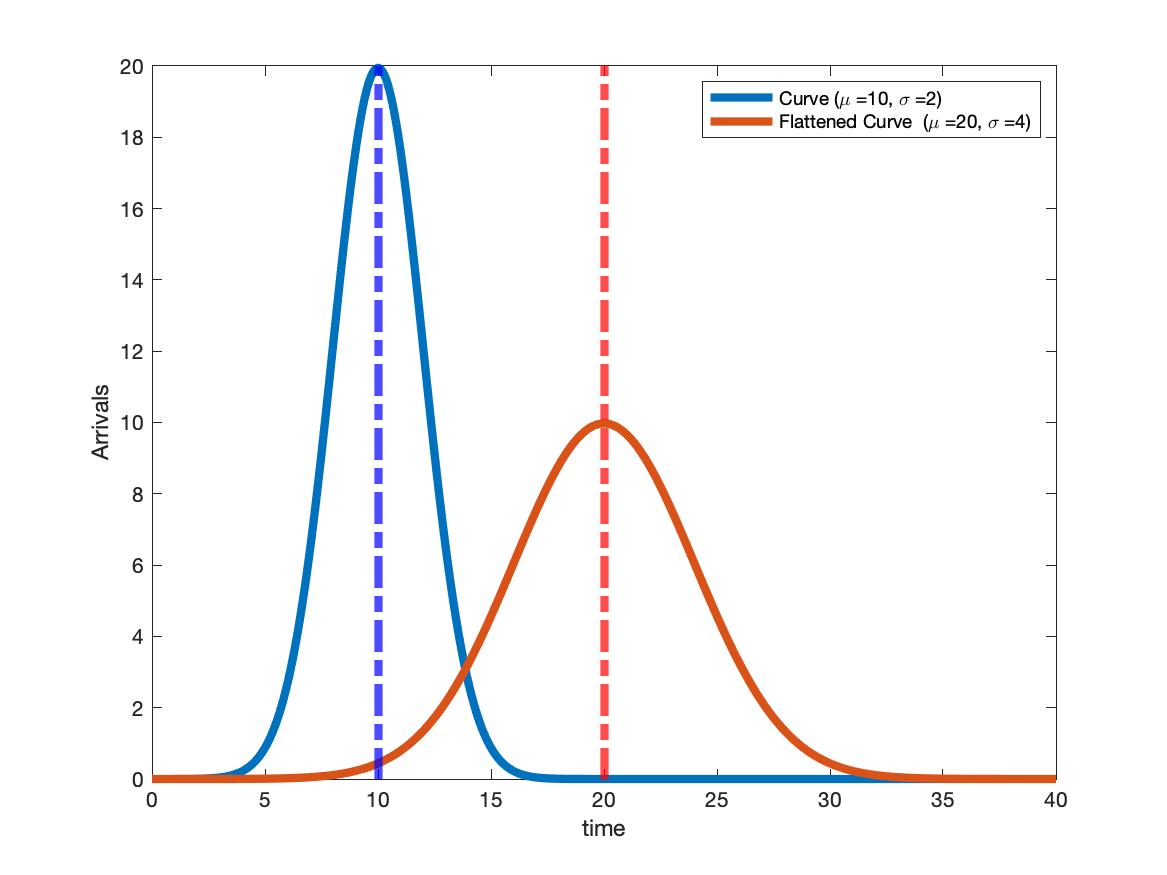}
	\centering
	\caption{Plot of Gaussian arrival rate functions.} \label{normal_fig}
\end{figure}

Using the Gaussian pdf as our arrival rate gives us the following integral relationships.

%\begin{equation}
%\int_{- \infty }^\infty  \lambda (t)dt = \lambda, \quad
%%\end{equation}
%%
%%\begin{equation}
%\frac{1}{\lambda }\int\limits_{ - \infty }^\infty  t\lambda (t)dt  = \tau, \quad
%%\end{equation}
%%
%%\begin{equation}
%\sqrt {\frac{1}{\lambda }\int\limits_{ - \infty }^\infty  {{{\left( {t - \tau } \right)}^2} \cdot \lambda (t)dt} }  = \sigma .
%\end{equation}

\begin{equation}
\int_{- \infty }^\infty  \lambda (t)dt = \lambda^*, \quad \frac{1}{\lambda^* }\int\limits_{ - \infty }^\infty  t\lambda (t)dt  = \tau, \quad \sqrt {\frac{1}{\lambda^* }\int\limits_{ - \infty }^\infty  {{{\left( {t - \tau } \right)}^2} \cdot \lambda (t)dt} }  = \sigma .
\end{equation}

%Using the closed form formulas for the $M_t/G/\infty$ queueing model from \citet{EMW}, we have for the time varying infinite server queue with Gaussian arrival rate that 
%\begin{eqnarray}
%{q_{M/G/\infty }}(t) &=& \frac{\lambda }{\sigma } \cdot {\text{E}}\left[ {\varphi \left( {\frac{{t - \tau  - {S_{\text{e}}}}}{\sigma }} \right)} \right] \cdot {\text{E}}S \\
%& \leq& \frac{{\lambda  \cdot {\text{E}}S}}{{\sigma \sqrt {2\pi } }}
%\end{eqnarray}  

In Figure \ref{normal_fig}, we plot two Gaussian arrival rate functions.  The blue curve is a Gaussian($\tau = 10, \sigma = 4$) and the red curve (flattened curve) is a Gaussian($\tau = 20, \sigma = 4$).  By increasing the standard deviation by a factor of 2, we have reduced the peak value of the arrival rate by a factor of two.  This follows from the mode of Gaussian being inversely proportional to the standard deviation $\sigma$.  Now that we have a full understanding of the Gaussian arrival rate function that we will use for patient arrivals, we can leverage insights from non-stationary queues to understand the dynamics of the total number infected and how long they stay infected.  

\begin{theorem} \label{queue_thm}
The $M_t/G/\infty$ queueing model with a Gaussian distribution arrival rate has a Poisson$(q_{M/G/\infty }(t))$ distribution where the mean $q_{M/G/\infty }(t)$ is the solution to the following ordinary differential equation 
\begin{eqnarray} \label{diffeqn}
 {\mathop q\limits^ \bullet}_{M/G/\infty }  (t) &=&  - \frac{\lambda^* }{\sigma ^2} \cdot E\left[ \left( \frac{t - \tau  - S_e}{ \sigma } \right) \varphi \left( \frac{t - \tau  - S_e}{\sigma } \right) \right] \cdot E[S]
\end{eqnarray}
and the solution is given by
\begin{eqnarray} \label{eqn_soln}
{q_{M/G/\infty }}(t) &=& \frac{\lambda^* }{\sigma } \cdot {\text{E}}\left[ {\varphi \left( {\frac{{t - \tau  - {S_{\text{e}}}}}{\sigma }} \right)} \right] \cdot E[S].
\end{eqnarray}  
Moreover, for any value of $t$, we have that 
\begin{eqnarray}
{q_{M/G/\infty }}(t) & \leq& \frac{{\lambda^*  \cdot E[S]}}{{\sigma \sqrt {2\pi } }}.  
\end{eqnarray}  
\begin{proof}
We actually start with the solution of the queue length given in Equation \ref{eqn_soln}.  The solution is easily given by the formula from \citet{EMW} since we have that 
\begin{eqnarray} \label{eqn_soln}
{q_{M/G/\infty }}(t) &=& \text{E} \left[ \lambda( t - S_e)  \right] \cdot E[S]\\
&=& \frac{\lambda^* }{\sigma } \cdot {\text{E}}\left[ {\varphi \left( {\frac{{t - \tau  - {S_{\text{e}}}}}{\sigma }} \right)} \right] \cdot E[S].
\end{eqnarray}  To find the differential equation, one simply takes the derivative of the solution with respect to the time parameter $t$.  The reason that we get the final expression in Equation \ref{eqn_soln} is that the derivative of the Gaussian distribution yields the first Hermite polynomial times the Gaussian pdf.  Finally, the bound on the standard Gaussian density function yields the bounds on the queue length.  This completes the proof.
\end{proof}
\end{theorem}

\begin{figure}[H] 
	\includegraphics[scale =.25]{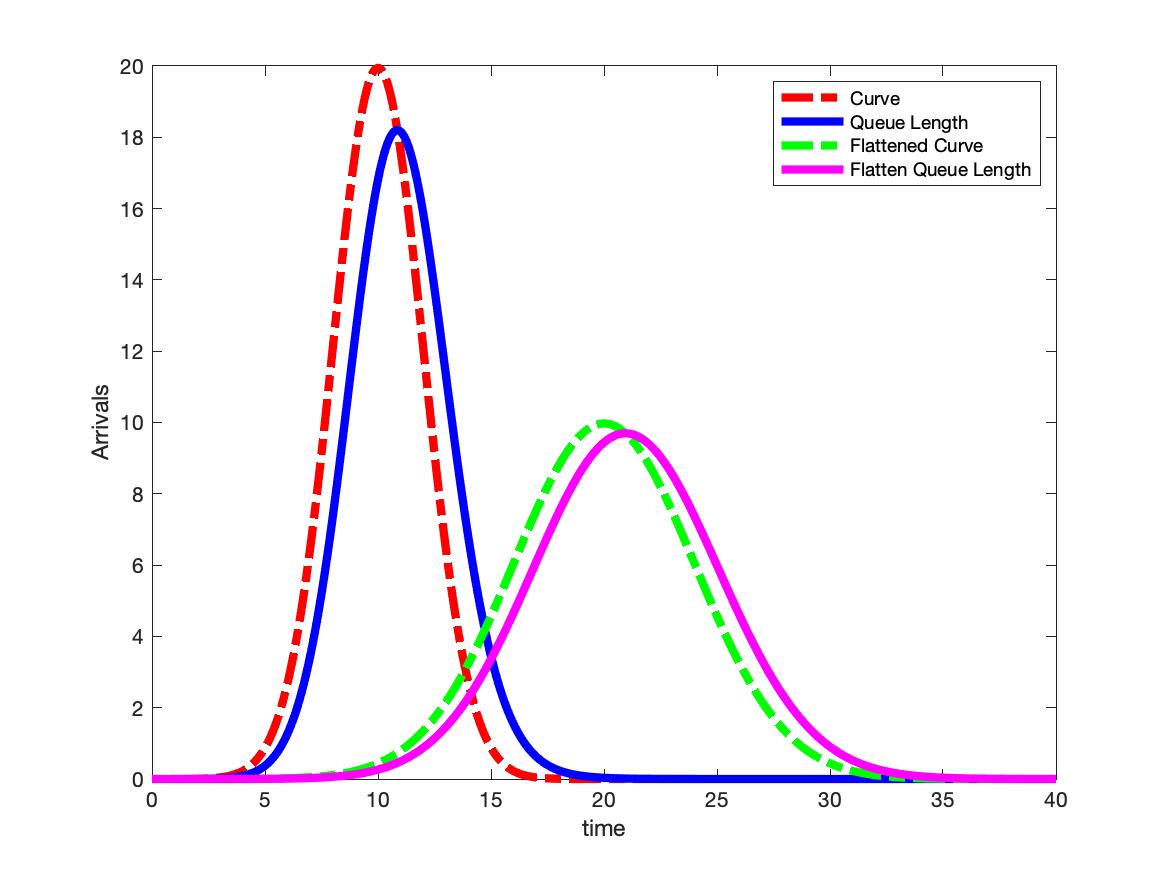}
	\centering
	\captionsetup{justification=centering}
	\caption{Gaussian arrival curves and their queue lengths ($\mathbb{E}[S] = 1, \lambda^* =100$).} \label{normal_queue}
\end{figure}

\begin{table}[H]
  \begin{center}
    \label{tab:table1}
    \begin{tabular}{|l|c|r|c|c|c|} % <-- Alignments: 1st column left, 2nd middle and 3rd right, with vertical lines in between
   \hline     \textbf{Curve Type} &  $\mathbf{\tau}$ &  $\mathbf{\sigma}$  & E[S] & \textbf{Peak Time} & \textbf{Peak Value}\\   \hline
      Arrival Curve & 10 &  2 & 1 & 10 & 19.95 \\  \hline
      Flattened Arrival Curve & 20 &  4 & 1 & 20 & 9.97 \\   \hline
      Queue Length  & 10&  2 & 1 &10.86 & 18.20  \\   \hline
      Flattened Queue Length  & 20 &  4 & 1 &  20.95 & 9.70  \\   \hline
    \end{tabular}
        \caption{Comparison between regular and flattened curves peak values and times ($\lambda^* =100$). } \label{table_1}
  \end{center} 
\end{table}

In Figure \ref{normal_queue}, we plot the same two Gaussian arrival functions given in Figure \ref{normal_fig}.  However, to the right of each arrival curve, we also plot the subsequent queue length when an exponential distribution is assumed for the service distribution.  The parameters for each curve and the service distributions are given in Table \ref{table_1}.  Table \ref{table_1} and Figure \ref{normal_queue} yield many observations.  

The first observation is that we see that both arrival rate curves intersect with the queue length at the peak queue length.  This is a nice property and can be explained by first noticing that $E[S] = 1/\mu  = 1$ in our example.  Thus, at the peak queue length, we know the time derivative of the queue length is equal to zero i.e.

\begin{eqnarray}
\shortdot{q}(t) &=& \lambda(t) - \mu q(t) = 0.
\end{eqnarray} 

This implies that at the time of the peak queue length, $t^*$ we have 
\begin{eqnarray} \label{peakk}
\lambda(t^*) &=& \mu q(t^*) .
\end{eqnarray} 

A second observation is that for this value of the service distribution, $E[S] =1$, the peak arrival rate is larger than the peak queue length.  We find that for the flattened arrival curve, the value of the peak queue length is closer to the value of  the peak arrival rate than in the non-flattened case.  This implies that even though one might flatten the curve and reduce the peak arrival rate by one half, it does not mean that the peak queue length are also reduced by one half.  We observe that the queue peak queue length is 91\% of the peak arrival rate in the non-flattened curve, while it is 97\% of the flattened curve.

In addition to finding the solution of the queue length, we can compute an expression for the time of the peak queue length.  We know that the time of the peak arrival rate occurs at the mode $\tau$ and our goal is to understand the lag effect i.e. the length of time the peak queue length occurs after the peak arrival rate occurs.  Theorem \ref{time_peak} suggests that the time of the peak queue length occurs after the peak arrival time $\tau$.  This is generally well known in the queueing literature.  However, this is an important fact for the COVID-19 outbreak as it helps determine exactly from the peak number of infections, when the peak load will be seen in the healthcare system.  

\begin{theorem} \label{time_peak}
The time of the peak queue length is the solution to the following fixed point equation
\begin{eqnarray}
t^* &=& \tau  + \frac{{{\text{E}}\left[ {{S_{\text{e}}} \cdot \varphi \left( {\frac{{t^* - \tau  - {S_{\text{e}}}}}{\sigma }} \right)} \right]}}{{{\text{E}}\left[ {\varphi \left( {\frac{{t^* - \tau  - {S_{\text{e}}}}}{\sigma }} \right)} \right]}} =  \tau  + \frac{{\lambda^*  \cdot {\text{E}}[S] \cdot {\text{E}}\left[ {{S_{\text{e}}} \cdot \varphi \left( {\frac{{t^* - \tau  - {S_{\text{e}}}}}{\sigma }} \right)} \right]}}{{\sigma  \cdot {q_{M/G/\infty }}\left( {t^*} \right)}}.
\end{eqnarray}
\begin{proof}
The proof is found in the Appendix.
\end{proof}
\end{theorem}

\begin{figure}[H] 
	\includegraphics[scale =.2]{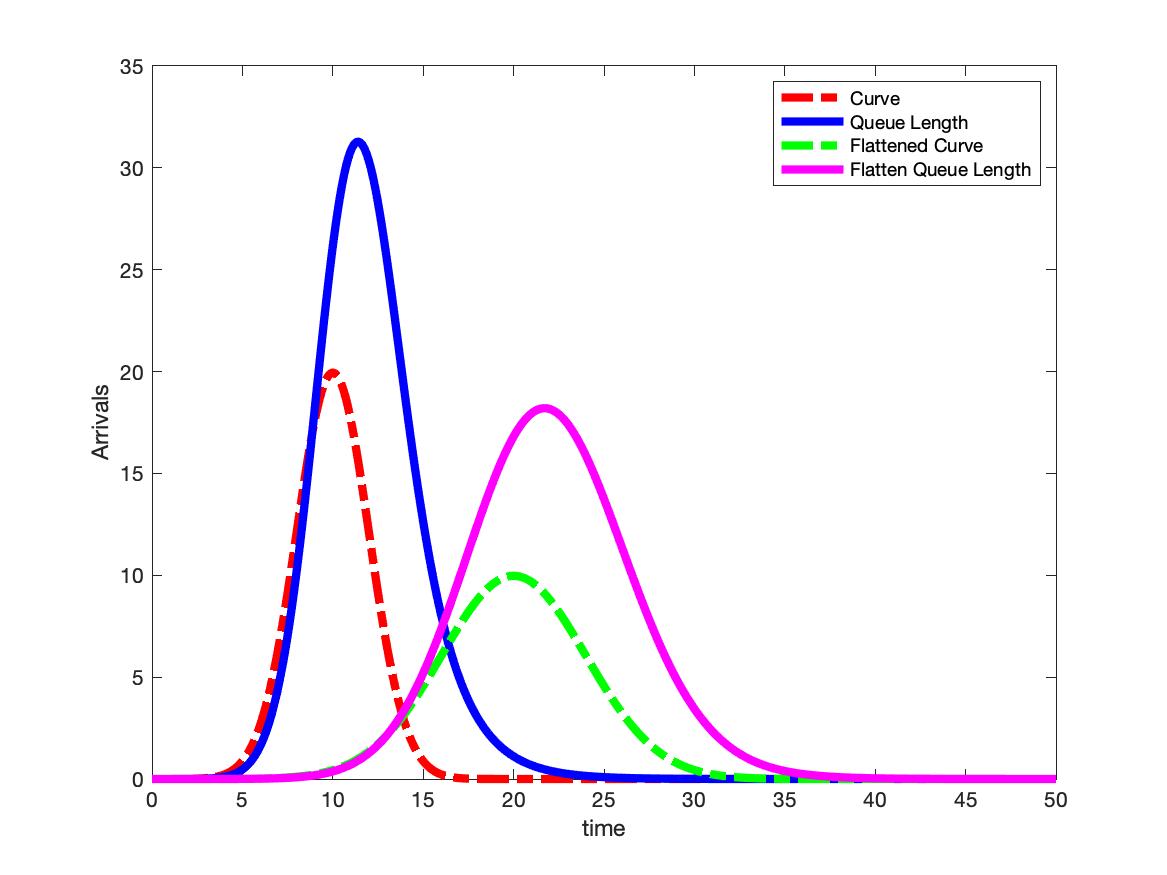}~\includegraphics[scale =.2]{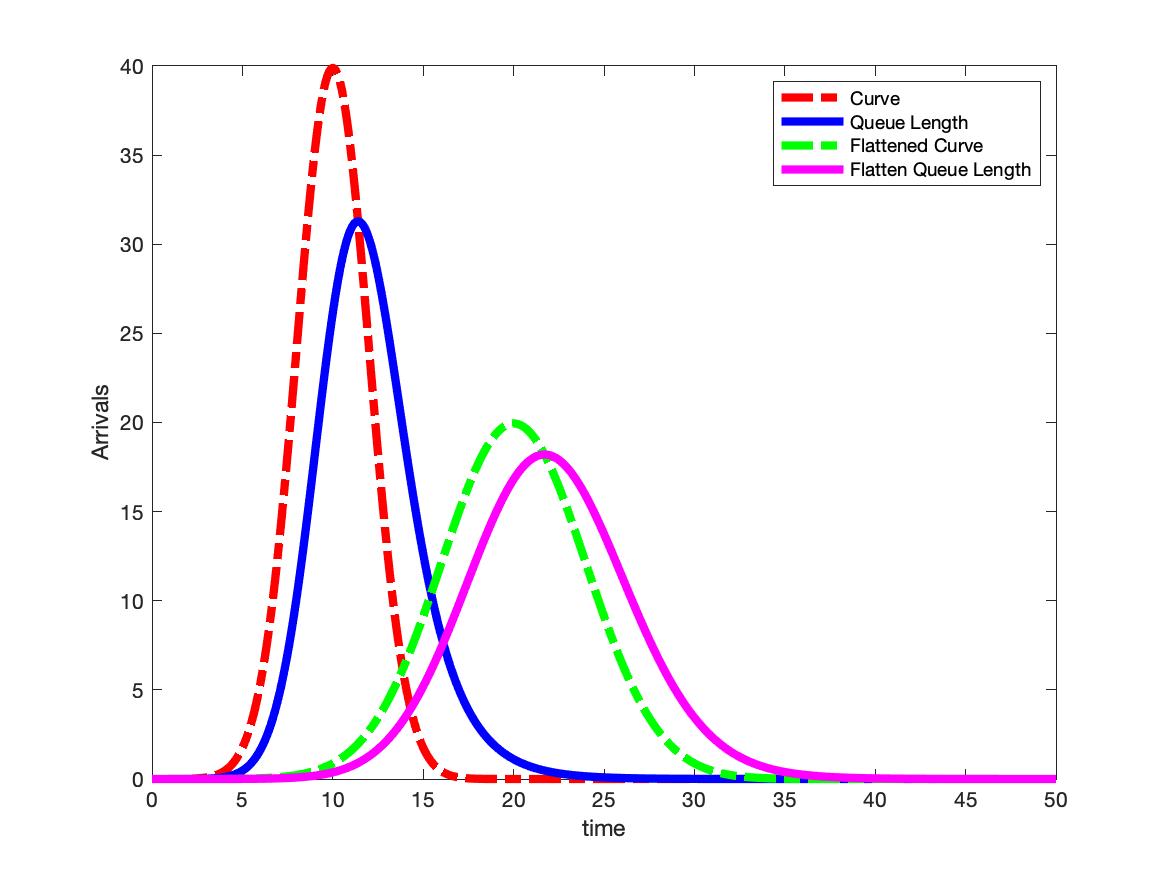}
	\centering
	\captionsetup{justification=centering}
	\caption{Gaussian arrival curves and their queue lengths ($\mathbb{E}[S] = 2, \lambda^* =100$).    \\ Unscaled arrival rate (Left).  Scaled Arrival rate (Right).} \label{normal_queue_2}
\end{figure}

\begin{table}[H]
  \begin{center}
    \label{tab:table1}
    \begin{tabular}{|l|c|r|c|c|c|} % <-- Alignments: 1st column left, 2nd middle and 3rd right, with vertical lines in between
   \hline     \textbf{Curve Type} &  $\mathbf{\tau}$ &  $\mathbf{\sigma}$  & E[S] & \textbf{Peak Time} & \textbf{Peak Value}\\   \hline
       Arrival Curve & 10 &  2 & 2 & 10 & 19.95 \\  \hline
      Flattened Arrival Curve & 20 &  4 & 2 & 20 & 9.97 \\   \hline
      Queue Length  & 10&  2 & 2 & 11.40 & 31.29  \\   \hline
      Flattened Queue Length  & 20 &  4 & 2 & 21.71 & 18.20  \\   \hline
    \end{tabular}
        \caption{Comparison between regular and flattened curves peak values and times ($\lambda^* =100$). } \label{table_2}
  \end{center}
\end{table}

\begin{figure}[H] 
\includegraphics[scale =.2]{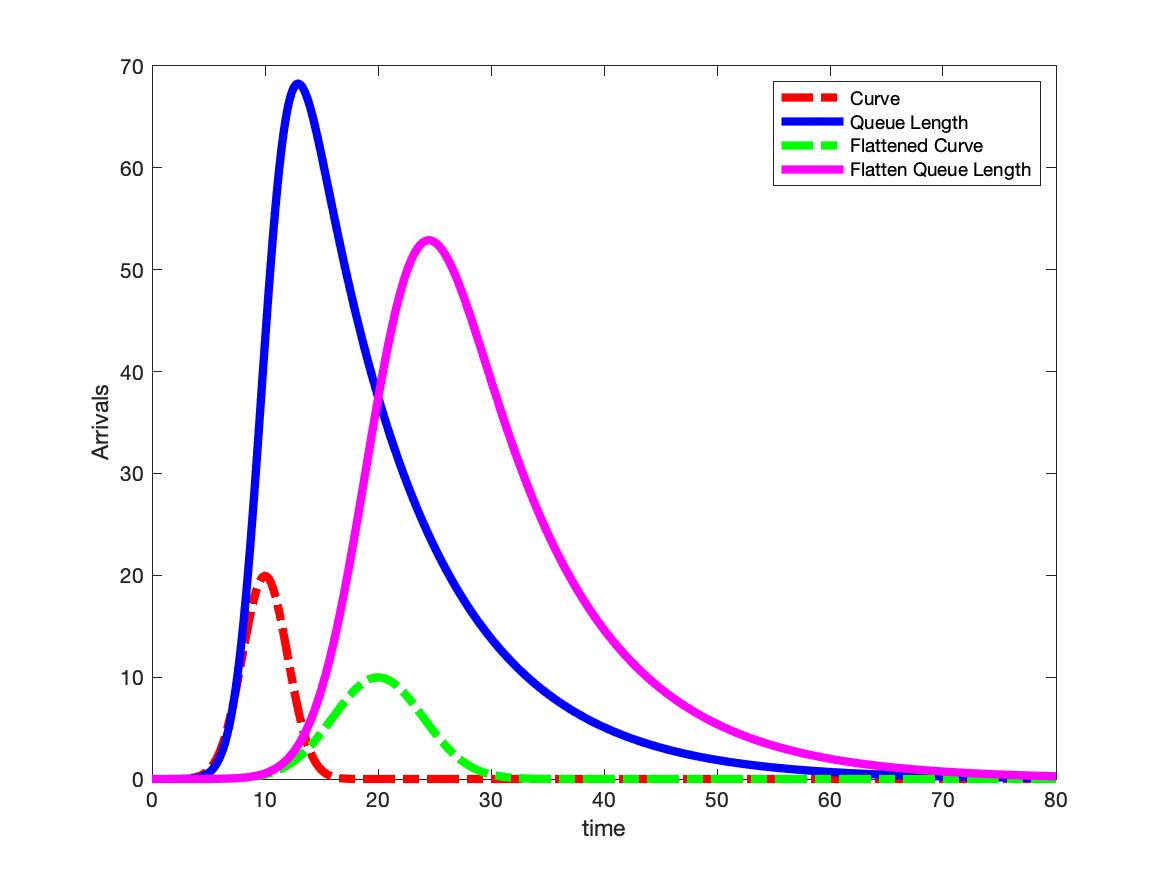}~\includegraphics[scale =.2]{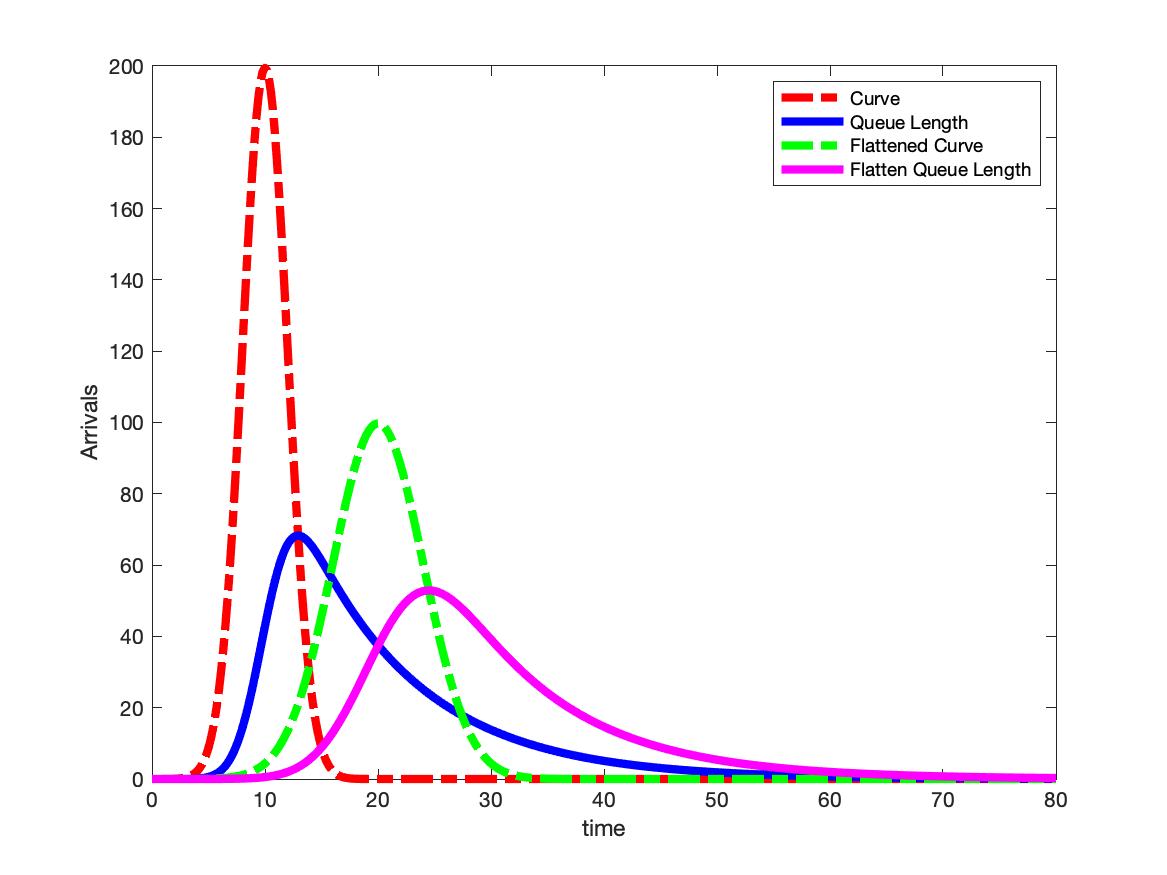}
	\centering
	\captionsetup{justification=centering}
	\caption{$\lambda^* = 100, \mu = .1, \sigma = 2, \sigma = 4, \tau = 10, \tau = 20$.} \label{normal_queue_3}
\end{figure}

\begin{table}[H]
  \begin{center}
    \label{tab:table1}
    \begin{tabular}{|l|c|r|c|c|c|} % <-- Alignments: 1st column left, 2nd middle and 3rd right, with vertical lines in between
   \hline     \textbf{Curve Type} &  $\mathbf{\tau}$ &  $\mathbf{\sigma}$  & E[S] & \textbf{Peak Time} & \textbf{Peak Value}\\   \hline
      Arrival Curve & 10 &  2 & 10 & 10 & 19.95 \\  \hline
      Flattened Arrival Curve & 20 &  4 & 10 & 20 & 9.97 \\   \hline
      Queue Length  & 10&  2 & 10 & 12.93 & 68.28  \\   \hline
      Flattened Queue Length  & 20 &  4 & 10 & 24.51 & 52.90  \\   \hline
    \end{tabular}
        \caption{Comparison between regular and flattened curves peak values and times ($\lambda^* =100$). }. \label{table_3}
  \end{center}
\end{table}

In Figures \ref{normal_queue_2} - \ref{normal_queue_3}, we plot our Gaussian arrival curves with their respective queue lengths.  In Figure \ref{normal_queue_2}, we set the mean service time to be $E[S] = 2$ and in Figure \ref{normal_queue_3}, we set the mean service time to be $E[S]= 10$.  On the left of the both Figures \ref{normal_queue_2} - \ref{normal_queue_3}, we plot the actual arrival rate and the actual queue length; however ,on the right of Figures \ref{normal_queue_2} - \ref{normal_queue_3}, we plot the actual queue length, but an arrival rate that is scaled by the mean service time $E[S]$.  We observe that the in the unscaled plots on the left, the arrival rate does not intersect with the queue length at the peak.  However, in the scaled plots on the right, we observe that the arrival rate intersects with the queue length curve at the queue length peak.  This verifies Equation \ref{peakk} again. Finally, we observe in Table \ref{table_1} that by reducing the arrival rate peak by one half reduces the queue length peak by one half when $E[S] =1$.  However,  we observe in Tables \ref{table_2} - \ref{table_3} that as we increase $E[S]$, we see a smaller reduction in the peak queue length.  Certainly we expect the infectious period to be longer than one day for COVID-19 so our theoretical results imply that we may need to do more flattening of the curve as patients spend more time infected.

%\begin{enumerate}
%\item Reducing arrival rate peak by one half reduces the queue length peak by one half when E[S] =1.  However, when $E[S] > 1$, we have a smaller reduction.  This is probably true of the real epidemic as people are infected for more than one day.  
%\item Figure \ref{normal_queue_2} shows this difference in impact of peak queue length.  
%\item Note how peak does not intersect with peak queue length, only in $E[S]=1$ case.  
%\end{enumerate}

\begin{theorem} \label{peak_bound}
The time of peak queue length is bounded by the following upper and lower bounds 
\begin{eqnarray}
\tau \leq t^* \leq  \tau  + \frac{E[S_e] }{\sqrt{2 \pi} \sigma {{\text{E}}\left[ {\varphi \left( {\frac{{t^* - \tau  - {S_{\text{e}}}}}{\sigma }} \right)} \right]}} = \tau  + \frac{E[S^2] }{\sqrt{8 \pi} \sigma E[S] {{\text{E}}\left[ {\varphi \left( {\frac{{t^* - \tau  - {S_{\text{e}}}}}{\sigma }} \right)} \right]}} .
\end{eqnarray}
\begin{proof}
First we observe that from Theorem \ref{time_peak} that $t^*$ solves the following fixed point equation
\begin{eqnarray}
t^* &=& \tau  + \frac{{{\text{E}}\left[ {{S_{\text{e}}} \cdot \varphi \left( {\frac{{t^* - \tau  - {S_{\text{e}}}}}{\sigma }} \right)} \right]}}{{{\text{E}}\left[ {\varphi \left( {\frac{{t^* - \tau  - {S_{\text{e}}}}}{\sigma }} \right)} \right]}} .
\end{eqnarray}
Now to prove the upper bound on the peak time, one observes that the numerator is bounded by the standard Gaussian bound given in Equation \ref{norm_bound}.  For the lower bound, one just recognizes that both expectations $ E\left[ {{S_{\text{e}}} \cdot \varphi \left( {\frac{{t^* - \tau  - {S_{\text{e}}}}}{\sigma }} \right)} \right] $ and $ E\left[ {\varphi \left( {\frac{{t^* - \tau  - {S_{\text{e}}}}}{\sigma }} \right)} \right] $  are non-negative.  To make an equivalence with the last equality, we use the equality between the expected stationary excess random variable and the second moment of the service distribution i.e. $E[S_e] = \frac{E[S^2]}{2 E[S]}$.
This completes the proof.  
\end{proof}
\end{theorem}

Theorem \ref{peak_bound} provides upper and lower bounds on the time of the peak queue length.  Numerically, one can use a bisection algorithm in order to explicitly determine the time of peak infection.  Once the time of the peak infection is known, then one can use Theorem \ref{queue_thm} to determine the queue length value at that peak time as well as the value of the peak infection.   This provides estimates on the number of ventilators or beds needed at the peak time of infection.

\begin{theorem}
Let ${q_{M/G/\infty }}(t)$ be a $M_t/G/\infty$ queueing model with a Gaussian distribution arrival rate.  If C is the maximum capacity queue length for all values of $t$, then the value of $\sigma^*$ that flattens the peak queue length below C is given by
\begin{eqnarray}
\sigma^* = \frac{{\lambda^*  \cdot {\text{E}}[S]}}{{ C \sqrt {2\pi } }} .
\end{eqnarray}     
\begin{proof}
We know the queue length is bounded by the constant 
\begin{eqnarray}
{q_{M/G/\infty }}(t) & \leq& \frac{{\lambda^*  \cdot {\text{E}}[S]}}{{\sigma \sqrt {2\pi } }}.  
\end{eqnarray}  
Now we let the constant equal the capacity and solve for the standard deviation to get 
\begin{eqnarray}
C & = & \frac{{\lambda^*  \cdot {\text{E}}[S]}}{{\sigma \sqrt {2\pi } }} \\
&\Rightarrow& \sigma^* = \frac{{\lambda^*  \cdot {\text{E}}[S]}}{{ C \sqrt {2\pi } }} .
\end{eqnarray}  
\end{proof}
\end{theorem}

\subsubsection{Deterministic Service Distribution}

In this section, we specify the service distribution to be a constant.  Although the constant assumption for the service time is quite unrealistic, this will simplify many of the complicated equations and yield simple insights.  Our first result shows the queue length can be described by an ode and the solution is simply an integral of the arrival rate evaluated at the current time and the current time minus the constant service rate.  Finally, we also show that the lag between the peak arrival rate and peak load is equal to half of the service time.    
\begin{corollary}
The $M_t/D/\infty$ queueing model with a Gaussian distribution arrival rate is the solution to the following ordinary differential equation 
\begin{eqnarray} \label{diffeqn}
 \mathop q\limits^ \bullet  (t) &=&  \lambda (t) - \lambda(t-\Delta) \\
 &=& \frac{\lambda^* }{\sigma } \cdot \varphi \left( {\frac{{t - \tau }}{\sigma }} \right) - \frac{\lambda^* }{\sigma } \cdot \varphi \left( {\frac{{t - \tau - \Delta }}{\sigma }} \right) 
 \end{eqnarray}
and the solution is given by
\begin{eqnarray} \label{det_soln}
{q_{M/G/\infty }}(t) &=&  \int^{t}_{t-\Delta} \frac{\lambda^* }{\sigma } \cdot \varphi \left( {\frac{{s - \tau }}{\sigma }} \right) ds \\
 &=& \lambda^* \left( \mathrm{erf}\left( \frac{ t-\frac{\tau}{\sigma}  }{\sqrt{2}} \right) - \mathrm{erf}\left( \frac{ t - \Delta - \frac{\tau}{\sigma} }{\sqrt{2}} \right) \right).
\end{eqnarray}  
Finally, the time of the peak is equal to 
\begin{eqnarray} 
t^* &=& \tau + \frac{\Delta}{2} .
\end{eqnarray}
\begin{proof}
This follows from standard analysis and the symmetry of the Gaussian distribution.  
\end{proof}
\end{corollary}

\subsubsection{Exponential Service Distribution}

When the service time distribution is exponential, the mean queue length, $q_{\infty}(t)$, solves the autonomous ordinary differential equation 
\begin{equation}
\shortdot{q}_{\infty}(t) = \lambda(t) - \mu \cdot q_{\infty}(t).
\end{equation}

The subsequent ordinary differential equation is linear and has a closed form solution given by the following expression

\begin{equation}
q_{\infty}(t) = q_0 e^{- \mu t } + e^{-\mu t}\int^{t}_{0} \lambda(s) e^{\mu s} \mathrm{d}s.
\end{equation}

\begin{corollary}
The $M_t/M/\infty$ queue with arrival rate given by a Gaussian distribution with parameters has the following closed form expression for the mean transient queue length 

\begin{eqnarray}
q_{\infty}(t) &=& \lambda^*  \cdot e^{ - \mu t } \cdot {e^{{\mu ^2}{\sigma ^2}/2}} \cdot \Phi \left( {\frac{t }{\sigma } - \mu \sigma } \right) .
\end{eqnarray}
%where the error function ($\mathrm{erf}$) and the normal distribution cumulative distribution function $\Phi(x)$ are  defined as 
%$$\mathrm{erf}(x) = \int^{x}_{0} \frac{e^{-z^2/2}}{\sqrt{2\pi}} dz \quad and \quad \Phi(x) = \int^{x}_{-\infty} \frac{e^{-z^2/2}}{\sqrt{2\pi}} dz$$

\begin{proof}
\begin{eqnarray}
q_{\infty}(t) &=&  \mathbb{E} [ \lambda( t - S_e) ] \cdot E[S] \\
&=& \frac{\lambda^*}{\sigma} \cdot \mu \sigma  \cdot {e^{ - \mu t }} \cdot {e^{{\mu ^2}{\sigma ^2}/2}} \cdot \Phi \left( {\frac{t }{\sigma } - \mu \sigma } \right) \cdot E[S] \\
&=& \lambda^*  e^{ - \mu t } \cdot {e^{{\mu ^2}{\sigma ^2}/2}} \cdot \Phi \left( {\frac{t }{\sigma } - \mu \sigma } \right)  
\end{eqnarray}
\end{proof}

\end{corollary}

\begin{theorem} \label{exact-lag}
When the arrival rate is a Gaussian function and the service time is given by an exponential distribution, then the lag $\ell$ in the time between the peak arrival rate and the time of the peak queue length is given by the following expression
\begin{eqnarray}
 \ell \equiv t^* - \tau =  \sigma \cdot( \mu \sigma + \psi(\mu \sigma) )
\end{eqnarray}
where 
\begin{eqnarray}
\psi^{-1}(x) = \frac{\varphi(x)}{\Phi(x)} .
\end{eqnarray}
%or where $\psi$ satisfies the following equation from \citet{hampshire2003provisioning}
%\begin{eqnarray}
%x = \frac{\varphi(\psi(x))}{\Phi(\psi(x))} 
%\end{eqnarray}
\begin{proof}
The proof is given in the Appendix.
\end{proof}
\end{theorem}

Thus, for the exponential distribution, we have an explicit formula for the lag in terms of the $\psi(\cdot)$ function.  This function is related to conditional expectations of Gaussian random variables and is related to the inverse Mills ratio i.e 
\begin{equation}
\psi^{-1}(x) = \mathbb{E} \left[ X | X \leq x \right] =   \frac{\varphi(x)}{\Phi(x)} .
\end{equation}  In the context of operations research, the $\psi(\cdot)$ function has been analyzed in resource sharing settings in \citet{hampshire2003provisioning, hampshire2009dynamic}, as well as, in the context of using risk measures for staffing multi-server queueing systems in \citet{pender2016risk}.

\subsection{The $M_t/G/\infty$ Queue with a Gamma Arrival Rate}
In this section, we describe the dynamics of the  $M_t/G/\infty$ queue whose arrival rate is driven by a Gamma distribution function.   Thus, the arrival rate function $\lambda(t)$ is given by the following expression
\begin{eqnarray}
\lambda (t, \alpha, \beta) \equiv \lambda(t) \equiv \frac{ \lambda^* \beta^{\alpha}}{\Gamma(\alpha)} t^{\alpha -1} e^{-\beta t}, \quad t \geq 0
\end{eqnarray}  
where $\alpha$ is shape parameter of the Gamma distribution, $\beta$ is the rate parameter Gamma distribution, $\lambda^*$ is the parameter. 
Unlike the standard Gaussian, the Gamma distribution is not guaranteed to be bounded unless the parameter $\alpha \geq 1$.  Thus, for the remainder of this paper, we assume that $\alpha \geq 1$.  Under this condition, we have that
\begin{equation} \label{norm_bound}
\lambda \left( t \right) \equiv  \frac{ \lambda^* \beta^{\alpha}}{\Gamma(\alpha)} t^{\alpha -1} e^{-\beta t} \leq \lambda \left( \frac{\alpha -1}{\beta} \right)  =   \frac{ \lambda^* \beta }{\Gamma(\alpha)} \left( \alpha -1 \right) ^{\alpha -1} e^{-(\alpha -1)}.
\end{equation}
Using the Gamma pdf as our arrival rate gives us the following integral relationships
\begin{equation}
\int_{0 }^\infty  \lambda (t)dt = \lambda^*, \quad \frac{1}{\lambda^* }\int\limits_{ 0 }^\infty  t\lambda (t)dt  = \frac{\alpha}{\beta}, \quad \sqrt {\frac{1}{\lambda^* }\int\limits_{ 0 }^\infty  {{{\left( {t - \tau } \right)}^2} \cdot \lambda (t)dt} }  = \frac{\alpha}{\beta^2} .
\end{equation}

One reason for using the Gamma distribution is that it is not symmetric like the Gaussian distribution.  From some of the data that we see in practice, the arrival rate distribution does not appear to be symmetric and may have a heavy tail.  For example in Figure \ref{data}, we see that the tail of the data is heavier than the Gaussian distribution tail.  This fact of the data encourages us to explore the Gamma arrival rate function which is asymmetric.  

\begin{figure}[ht] 
	\includegraphics[scale =.23]{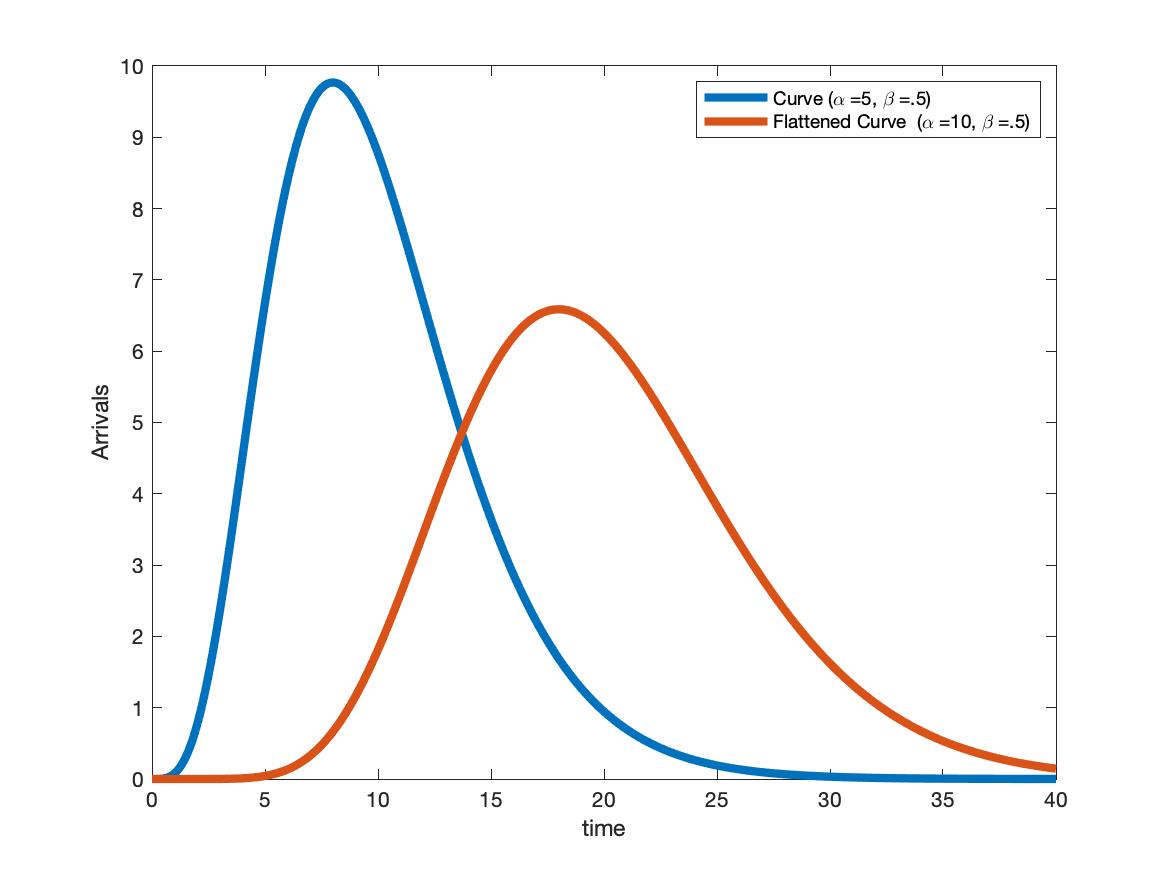}
	\centering
	\caption{Plot of Gamma Arrival Rate Curves.} \label{gamma_fig}
\end{figure}

In Figure \ref{gamma_fig}, we plot two gamma arrival rate functions.  The blue curve is a Gamma($\alpha = 5, \beta = .5$) and the red curve (flattened curve) is a Gamma($\alpha = 10, \sigma = .5$).  By increasing the variance by a factor of 2, we have reduced the peak value of the arrival rate by a factor of two.  This follows from the mode of gamma being inversely proportional to the standard deviation $\sigma$.  Now that we have a full understanding of the gamma arrival rate function that we will use for patient arrivals, we can leverage insights from non-stationary queues to understand the dynamics of the total number infected and how long they stay infected.  

%\begin{figure}[ht]
%	\includegraphics[scale =.35]{./Figures/XCHINAdailyInfectionRate.jpg}~\includegraphics[scale =.35]{./Figures/XSKOREAdailyInfectionRate.jpg}
%	\centering
%	\caption{China confirmed infection data (Left).  South Korea confirmed infection data (Right).} \label{gamma_fig}
%	 \label{data}
%\end{figure}  
%
%\begin{figure}[ht]
%	\includegraphics[scale =.35]{./Figures/XCHINAdailyDeathRate.jpg}~\includegraphics[scale =.35]{./Figures/XSKOREAdailyDeathRate.jpg}
%	\centering
%	\caption{China death data (Left).  South Korea death data (Right).} \label{gamma_fig}
%	\label{data}
%\end{figure}  

%
%
%\textit{Written by Sergio}
%\textit{The normal arrival rate function is fitted to the data by setting the mean, $\tau$ equal to the time in which the empirical arrival rate was greatest. $\lambda$ was set to equal the mean number of total infections as measured each day plus an explanation parameter $\epsilon$ that determines how thick the curve is about the mean. The $\epsilon$ parameter helps determine how conservative you want th model to be, where sufficiently large increases in $\epsilon$ increases the number of points under the curve. Having set $tau$, $\lambda$, and $\epsilon$ to specific values $\sigma$ is solved for. The gamma distribution parameter $\theta$ (the scale parameter) was set to $2$, the mode was set to the time of the maximal arrival rate from which the shape parameter $k$ was solved for.}

\begin{theorem} \label{queue_thm_gamma}
The $M_t/G/\infty$ queueing model with a Gamma$(\alpha, \beta)$ distribution arrival rate has a Poisson$(q_{M/G/\infty }(t))$ distribution where the mean $q_{M/G/\infty }(t)$ is the solution to the following ordinary differential equation 
\begin{eqnarray} \label{diffeqn_gamma}
 {\mathop q\limits^ \bullet}_{M/G/\infty }  (t) &=&  \frac{\lambda^* \beta^\alpha}{\Gamma(\alpha)} \cdot E\left[ \left( \alpha - 1 - \beta (t - S_e) \right) \cdot \left( (t   - S_e)^{\alpha -2} e^{-\beta (t - S_e)}  \right) \right] \cdot E[S]
\end{eqnarray}
and the solution is given by
\begin{eqnarray} \label{eqn_soln_gamma}
{q_{M/G/\infty }}(t) &=&\frac{\lambda^* \beta^\alpha}{\Gamma(\alpha)} \cdot E\left[ (t   - S_e)^{\alpha -1} e^{-\beta (t - S_e)}  \right] \cdot E[S].
\end{eqnarray}  
Moreover, for any value of $t$ and $\alpha \geq 1$, we have that 
\begin{eqnarray}
{q_{M/G/\infty }}(t) & \leq&  \frac{ \lambda^* \beta }{\Gamma(\alpha)} \left( \alpha -1 \right) ^{\alpha -1} e^{-(\alpha -1)} \cdot E[S].
\end{eqnarray}  
\begin{proof}
We actually start with the solution.  The solution is easily given by the the formula from \citet{EMW}.  To find the differential equation, one simply takes the derivative of the solution with respect to the time parameter $t$.  Finally, the bound on the standard Gaussian density function yields the bounds on the queue length.  
\end{proof}
\end{theorem}

\begin{theorem} \label{time_peak_gamma}
The time of the peak queue length is the solution to the following fixed point equation
\begin{eqnarray}
t^* &=& \frac{\alpha-1}{\beta}  + \frac{ \mathbb{E}\left[ S_e \cdot \left( (t^*   - S_e)^{\alpha -2} e^{-\beta (t^* - S_e)}  \right) \right] }{ \mathbb{E}\left[ (t^*   - S_e)^{\alpha -2} e^{-\beta (t^* - S_e)} \right] } .
\end{eqnarray}
\begin{proof}
In order to find the peak queue length we need to set the time derivative of the queue length to zero i.e.
\begin{equation}
\mathop q\limits^ \bullet  (t^*) = 0 .
\end{equation}
Now using the differential equation given in Equation \ref{diffeqn_gamma}, we have 
\begin{eqnarray}
\mathop q\limits^ \bullet  (t^*) =   \frac{\lambda^* \beta^\alpha}{\Gamma(\alpha)} \cdot E\left[ \left( \alpha - 1 - \beta (t - S_e) \right) \cdot \left( (t   - S_e)^{\alpha -2} e^{-\beta (t - S_e)}  \right) \right] \cdot E[S] = 0.
\end{eqnarray}

Isolating $t^*$ by itself on the left hand side we obtain
\begin{eqnarray}
t^* &=& \frac{\alpha-1}{\beta}  + \frac{ \mathbb{E}\left[ S_e \cdot \left( (t^*   - S_e)^{\alpha -2} e^{-\beta (t^* - S_e)}  \right) \right] }{ \mathbb{E}\left[ (t^*   - S_e)^{\alpha -2} e^{-\beta (t^* - S_e)} \right] } .
\end{eqnarray}
This completes the proof.  
\end{proof}
\end{theorem}

What is nice about this result is that it decomposes into two parts.  The first part is the mode of the arrival rate function, which in the case of the Gamma distribution occurs at $\frac{\alpha-1}{\beta}$.  The second part is the positive shift from the mode of the arrival rate function, which is $\frac{ \mathbb{E}\left[ S_e \cdot \left( (t^*   - S_e)^{\alpha -2} e^{-\beta (t^* - S_e)}  \right) \right] }{ \mathbb{E}\left[ (t^*   - S_e)^{\alpha -2} e^{-\beta (t^* - S_e)} \right] }$.  This also implies that the peak queue length occurs after the peak arrival rate like in the Gaussian case.    

\begin{figure}[H] \label{gamma_queue}
	\includegraphics[scale =.2]{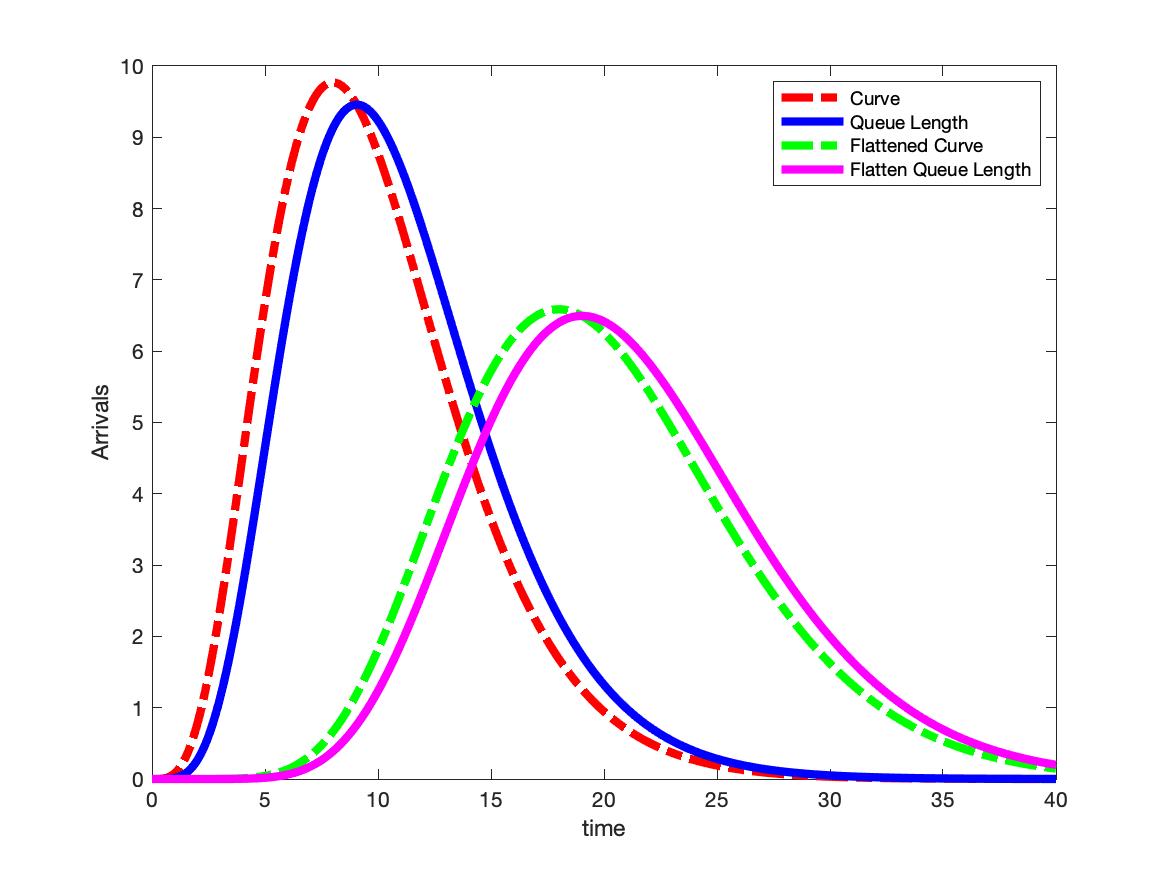}~\includegraphics[scale =.2]{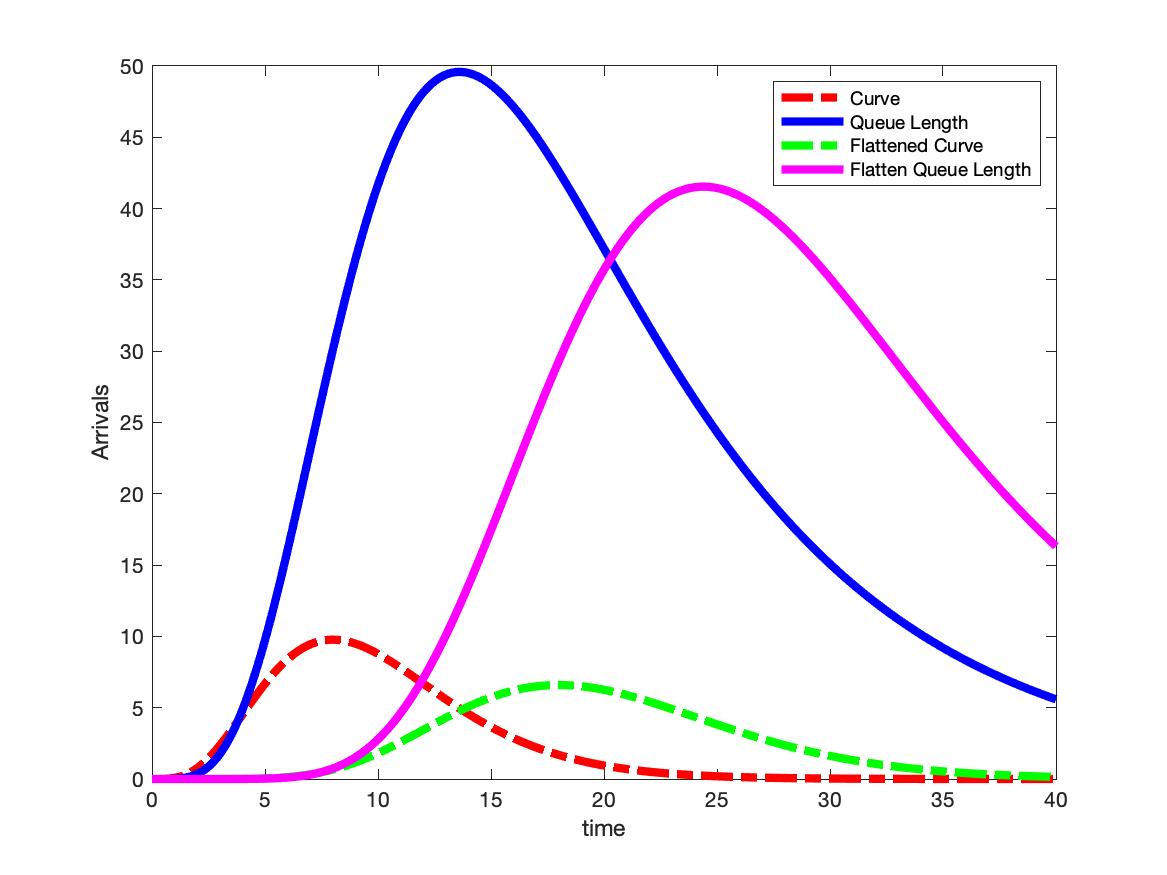}
	\centering
	\captionsetup{justification=centering}
	\caption{Gamma arrival curves and their queue lengths $E[S] = 2$ (Left) and $E[S] = 10$ (Right). }
\end{figure}

\begin{table}[H]
  \begin{center}
    \label{tab:table1}
    \begin{tabular}{|l|c|r|c|c|c|} % <-- Alignments: 1st column left, 2nd middle and 3rd right, with vertical lines in between
   \hline     \textbf{Curve Type} &  $\mathbf{\alpha}$ &  $\mathbf{\beta}$  & E[S] & \textbf{Peak Time} & \textbf{Peak Value}\\   \hline
      Gamma Arrival Curve & 5 &  2 & 1 & 8.00 & 9.77 \\  \hline
      Flattened Gamma Arrival Curve & 10 &  2 & 1 & 18.00 & 6.59 \\   \hline
      Queue Length  & 5&  2 & 1 &9.06 & 9.46 \\   \hline
      Flattened Queue Length  & 10 &  2 & 1 &  19.03 & 6.50  \\   \hline
    \end{tabular}
        \caption{Comparison between regular and flattened curves peak values and times.}
  \end{center}
\end{table}

%\begin{figure}[ht] \label{gamma_queue}
%	\includegraphics[scale =.4]{./Figures/Figure_24.jpg}
%	\centering
%	\caption{Data.}
%\end{figure}

\begin{table}[H]
  \begin{center}
    \label{tab:table1}
    \begin{tabular}{|l|c|r|c|c|c|} % <-- Alignments: 1st column left, 2nd middle and 3rd right, with vertical lines in between
   \hline     \textbf{Curve Type} &  $\mathbf{\alpha}$ &  $\mathbf{\beta}$  & E[S] & \textbf{Peak Time} & \textbf{Peak Value}\\   \hline
      Gamma Arrival Curve & 5 &  2 & 10 & 8.00 & 9.77 \\  \hline
      Flattened Gamma Arrival Curve & 10 &  2 & 10 & 18.00 & 6.59 \\   \hline
      Queue Length  & 5&  2 & 10 &13.60 & 49.59 \\   \hline
      Flattened Queue Length  & 10 &  2 & 10 &  24.39 & 41.54  \\   \hline
    \end{tabular}
        \caption{Comparison between regular and flattened curves peak values and times.}
  \end{center}
\end{table}

\section{Understanding COVID-19 Through Data}

\begin{figure}[H] \label{gamma_queue}
	\includegraphics[scale =.34]{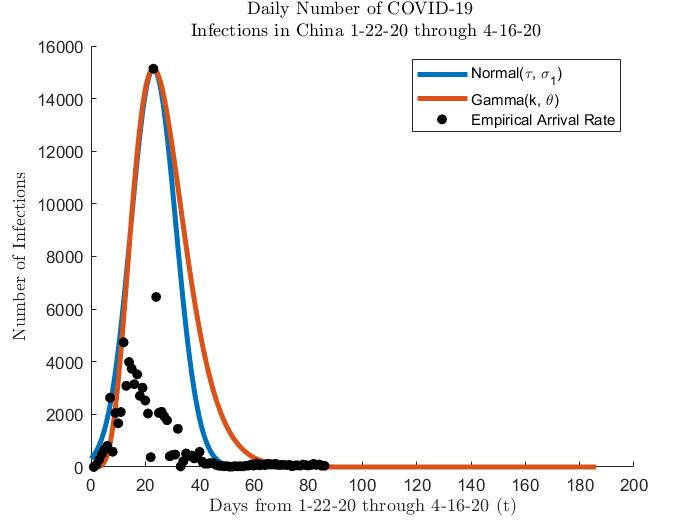}~\includegraphics[scale =.34]{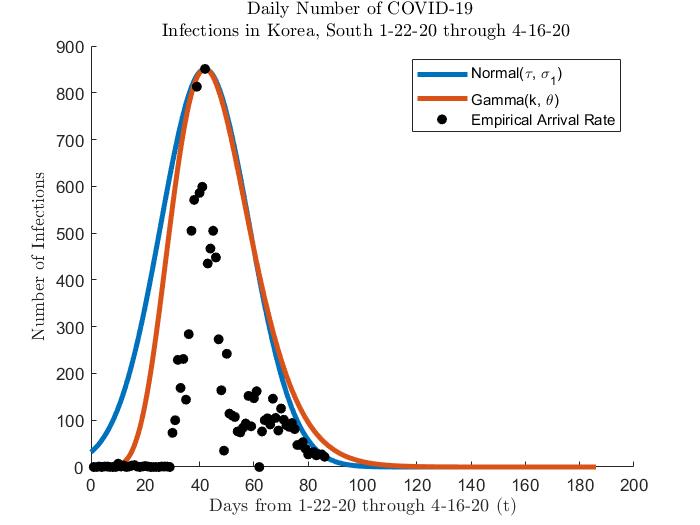}
	\includegraphics[scale =.34]{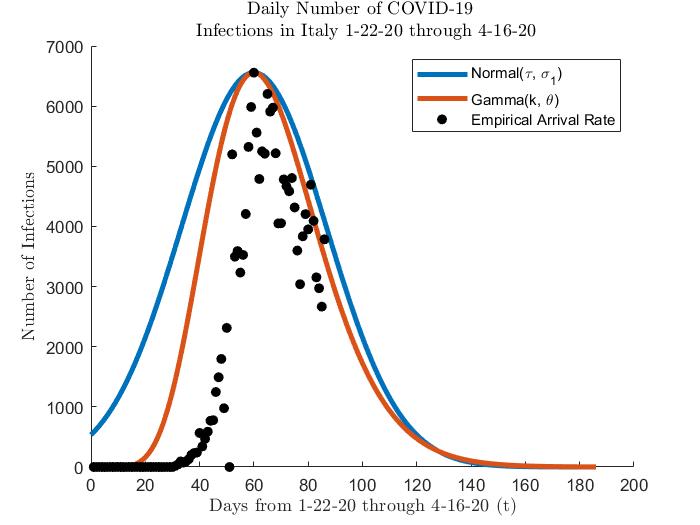}~\includegraphics[scale =.34]{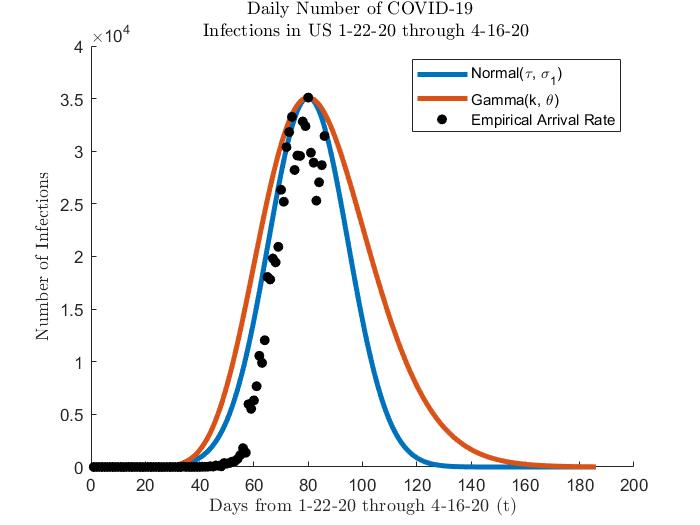}
	\centering
	\captionsetup{justification=centering}
	\vspace{-0in}
	\caption{Arrival Scatter Plots. } \label{arrivals}
\end{figure}

In this section, we use some of the data on COVID-19 infections and deaths, which was made available through Johns Hopkins University website \citet{dong2020interactive}.  In Figure \ref{arrivals}, we plot the number of confirmed infections per day in the countries of China, South Korea, Italy, and the United States.  One immediately can tell these countries apart since the confirmed cases start earlier in China, then South Korea, then Italy, and then finally the United States.  We also observe that as of April 16th, it appears that the United States has hit its peak and is on its way down.  On each of the plots, we also overlay a Gaussian and gamma distribution function to get a sense what distribution parameters best fit the data.  For China, the best Gaussian fit was a $Gaussian( \lambda^* =314101,  \tau = 23, \sigma = 8.28)$ and the best gamma fit was a $Gamma( \lambda^* =369218, \alpha = 6.75 , \beta = .25)$.  For South Korea, the best Gaussian fit was a $Gaussian( \lambda^* =34914,  \tau = 42, \sigma = 16.37)$ and the best gamma fit was a $Gamma( \lambda^* =31220, \alpha = 9.4 , \beta = .2)$.  For Italy, the best Gaussian fit was a $Gaussian( \lambda^* =440393,  \tau = 60, \sigma = 26.79)$ and the best gamma fit was a $Gamma( \lambda^* = 340126, \alpha = 9.57 , \beta = .15)$.  For the United States, the best Gaussian fit was a $Gaussian( \lambda^* = 1295233,  \tau = 80, \sigma = 14.72 )$ and the best gamma fit was a $Gamma( \lambda^* =1768739, \alpha = 17, \beta =.2)$.

\begin{figure}[H] \label{gamma_queue}
	\includegraphics[scale =.34]{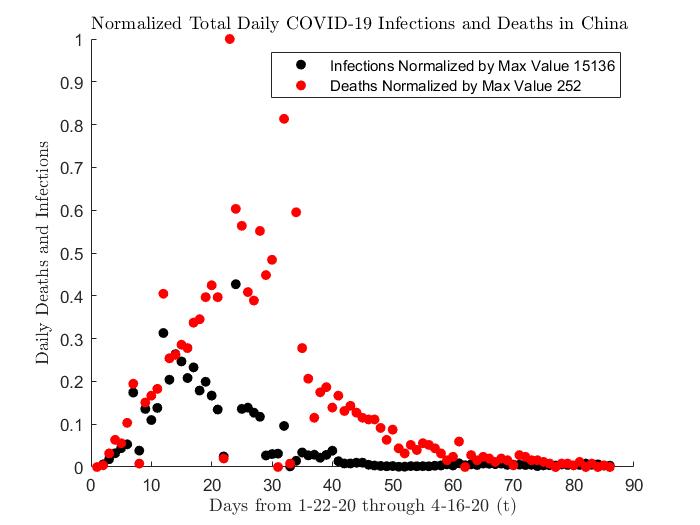}~\includegraphics[scale =.34]{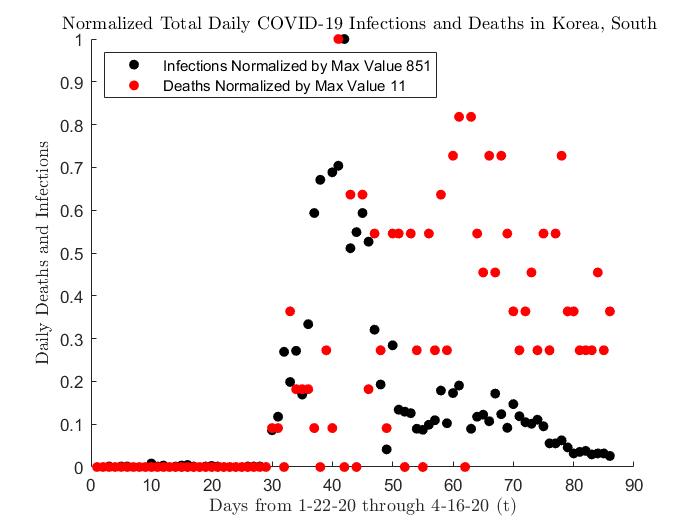}
	\vspace{2.2in}
	\includegraphics[scale =.34]{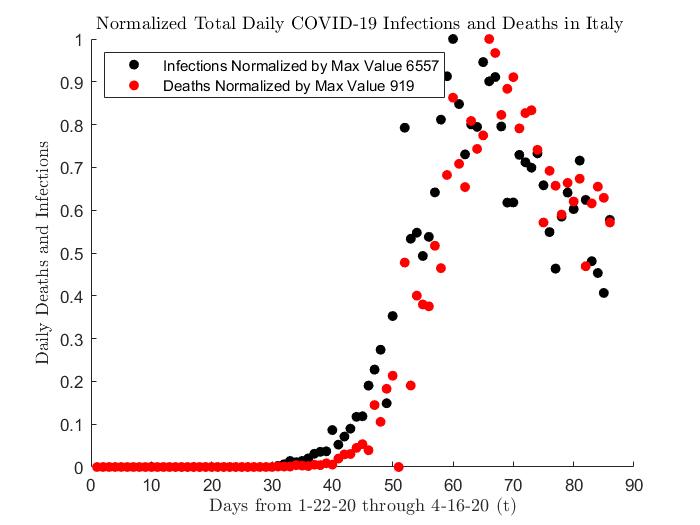}~\includegraphics[scale =.34]{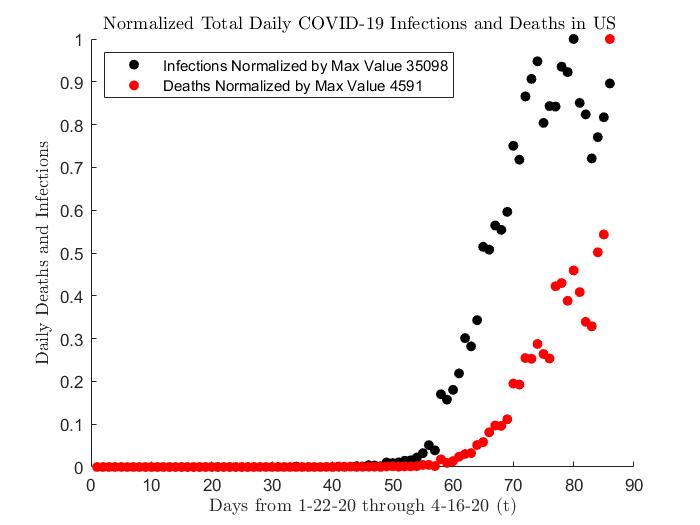}
	\centering
	\captionsetup{justification=centering}
	\vspace{-2in}
	\caption{Arrival and Death Scatter Plots. } \label{arrivals_deaths}
\end{figure}

In Figure \ref{arrivals_deaths}, we plot the arrivals and deaths per day from January 22nd to April 16th normalized by their maximum value (arrival = black dots, deaths = red dots).  For China, South Korea, Italy, and the United States, the maximum confirmed infections was (15,136), (851), (6557), and (35,098) respectively.  Moreover, for China, South Korea, Italy, and the United States, the maximum confirmed  deaths was (252), (11), (919), and (4591) respectively. We observe that for most all countries except Italy, the largest number of deaths lags the largest number of confirmed infections.  This is the lag effect that we observe in our queueing models.  This lag effect also inspires our construction of a new way to view the lag effect.

\begin{figure}[H] \label{gamma_queue}
	\includegraphics[scale =.34]{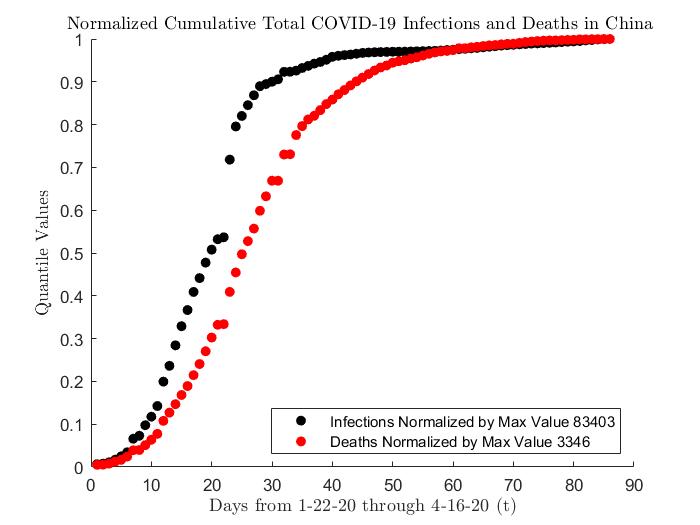}~\includegraphics[scale =.34]{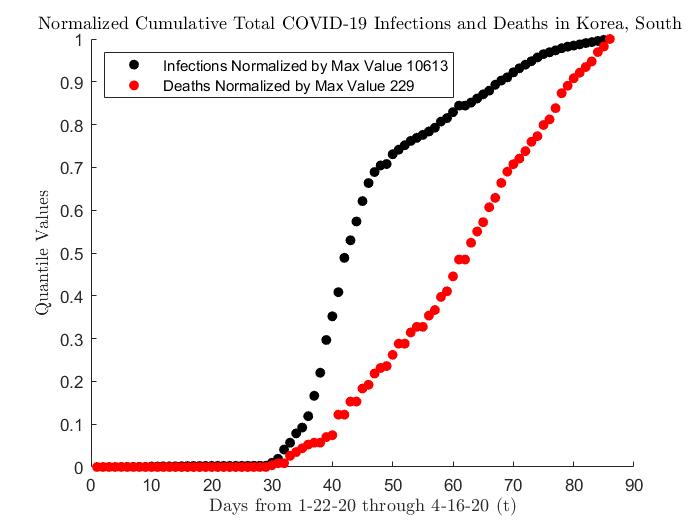}
	\includegraphics[scale =.34]{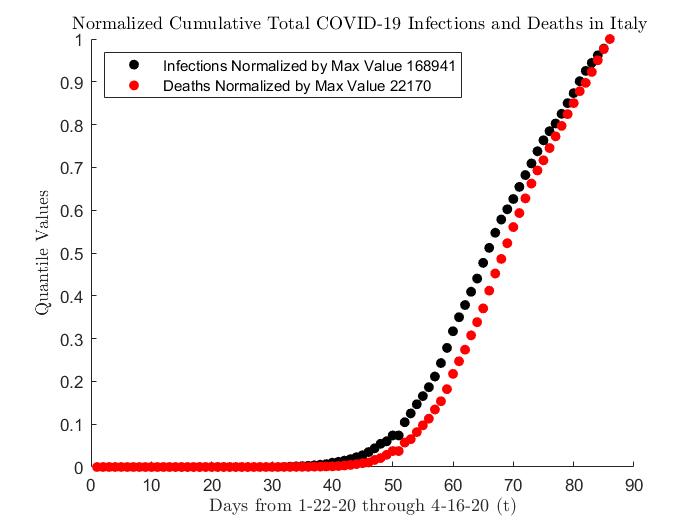}~\includegraphics[scale =.34]{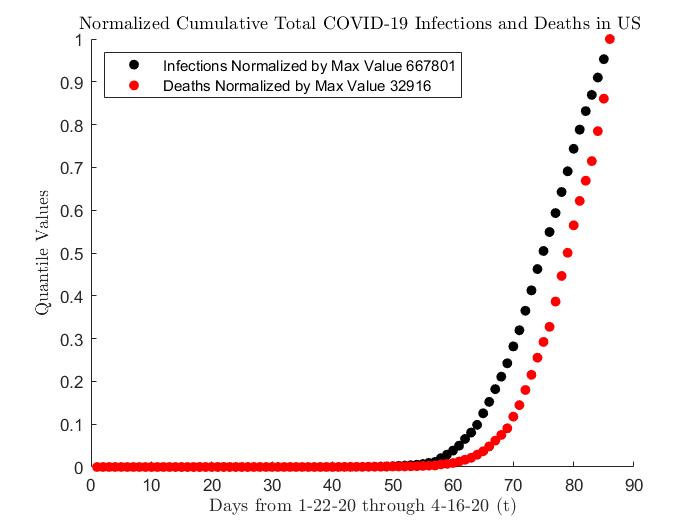}
	\includegraphics[scale =.34]{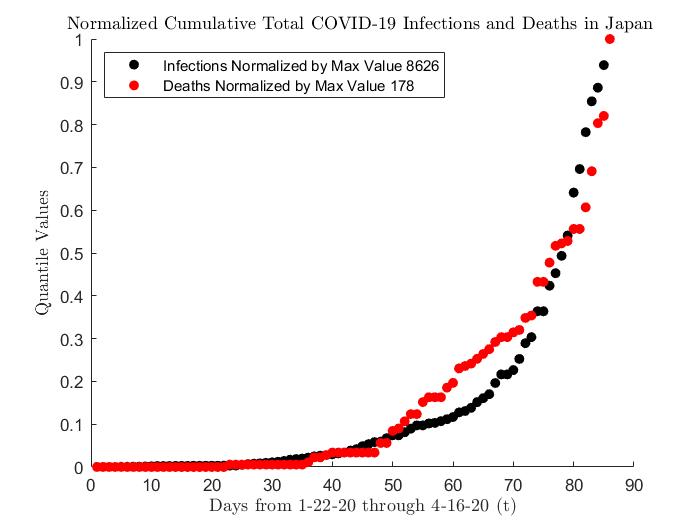}~\includegraphics[scale =.34]{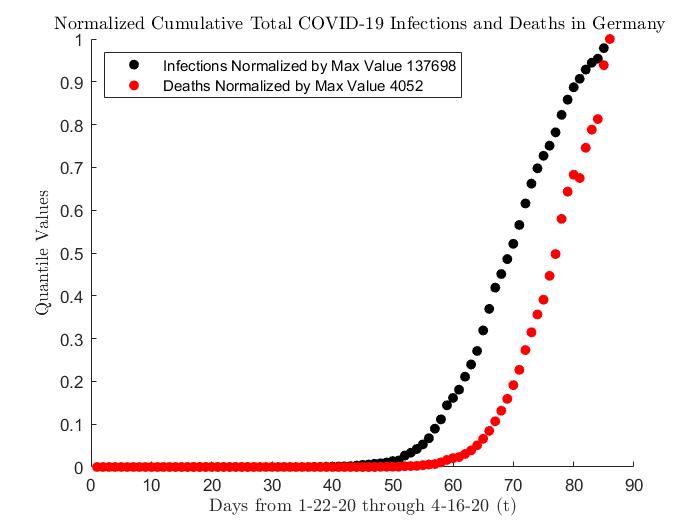}
	\centering
	\captionsetup{justification=centering}
	\caption{CDF of Arrival and Death Numbers. } \label{cdfs}
\end{figure}

In Figure \ref{cdfs}, we plot the cumulative distribution function of the total number of arrivals and deaths for the countries of China, South Korea, Italy, United States, Japan, and Germany.  This plot matches the quantiles of the arrivals to the quantiles to that of the deaths.  In this plot we observe for 5 out of the 6 countries that the cdf of the death count lags behind the arrival cdf.  The only country that does not follow the lag effect is Japan and this observation is consistent with Japan not testing its citizens enough \citet{Japantest}.  Moreover, we also observe that the lag effect is most pronounced in South Korea, which has done a great job of testing its citizens and keeping them alive \citet{Koreatest}.  If we rank the countries according to their lag effects, it is from best to worst (South Korea, Germany, China, United States, Italy, Japan).   This method of viewing the arrival and death cdfs allows one to understand the impact of testing and the quality of the healthcare system in that country.   These observations are also confirmed in Table \ref{tab:table1} where we plot the lag according to the quantile levels in the cdf plots given in Figure \ref{cdfs}.  

\begin{table}[H]
  \begin{center}
    \begin{tabular}{|l|c|c|c|c|c|c|c|c|c|} % <-- Alignments: 1st column left, 2nd middle and 3rd right, with vertical lines in between
   \hline     \textbf{Country} &  $q = .1$ &  $q = .2$  & q = .3& q = .4& q = .5 &q = .6 &  $q = .7$  & q = .8& q = .9\\   \hline
      China & 2 &  4 & 5 & 6 & 6 &  6 & 9 & 11 & 14  \\  \hline
      Germany& 9 &  9 & 8 & 9 & 8 &  7 & 7 & 6 & 4  \\  \hline
      Italy & 4 &  3 & 3 & 3 & 3 &  3 & 2 & 2 & 2  \\  \hline
      Japan & -4 &  -7 & -5 & -2 & -2 &  2 & 2 & 1 & 1  \\  \hline
      South Korea & 5 &  9 & 13 & 18 & 20 &  21 & 22 & 18 & 12  \\  \hline
      United States & 5 &  5 & 5 & 5 & 4 &  3 & 3 & 3 & 2  \\  \hline

    \end{tabular}
        \caption{Lags between arrival and death cdfs by quantile.}  \label{tab:table1}
  \end{center}
\end{table}

\section{Key Takeaways and Insights}

The analysis of the queueing model provides both qualitative insights and prescriptive guidance on how to flatten the demand for hospital resources.

\subsection{Flattening the Curve}
First, we show that the peak number of hospitalized patients is inversely related to flattening the rate of arriving patients (Theorem \ref{queue_thm}). This insight provides a mechanism through the standard deviation parameter $\sigma$ to control the peak number of hospitalized people.  However, the effectiveness of social distancing on flattening the curve, i.e. increasing $\sigma$, remains an open question. As a result, in order to avoid overcrowding flattening should be inversely proportional to the amount of hospital capacity. Additionally, we find that flattening the curve has a smaller impact on the peak number of patients when the service times are long.  This means the longer patients stay in the hospital, the more society must flatten the curve in order to achieve the same capacity level.  In the three examples with an exponential service distribution and a Gaussian arrival rate, we flattened the peak arrival rate by 50\%, however, we see that the peak queue length was reduced by 47\%, 42\%, 23\% respectively for E[S] = \{1,2,10\}.  For the gamma arrival rate setting, we see that a 33\% reduction in the peak arrival rate, we obtain a 31\% and 16\% reduction respectively for  E[S] = \{1,10\}.

\subsection{Lag between the Peak Infection Rate and the Peak Number of Patients}

A key takeaway from the queueing analysis is that the time of the peak arrival rate of infected patients does not coincide with the time of the peak of number of hospitalized people (or deaths). There is a time lag. We characterize this lag exactly for the simplified queue model (Theorem \ref{time_peak}). First, we show that the peak number of infected lags behind the peak number of new infected individuals.  Surprisingly, there is a nonlinear relationship between flattening the curve and the time lag between the peak rate of newly admitted patients and the peak demand for hospital resources (Theorem \ref{time_peak}).  Moreover, we show in Theorem \ref{peak_bounds} that the lag, $\ell$,  is bounded by the mean service time for an exponential distribution i.e.
\begin{eqnarray}
\frac{1}{ \mu} - \frac{1}{\mu^3 \sigma^2 + \mu } \leq \ell \leq \frac{1}{\mu} = E[S].
\end{eqnarray}  
This analysis also implies that as we flatten the curve i.e. increase $\sigma$, the lag converges to the mean service time $\frac{1}{\mu}$.

 \section{Conclusion}
 
In this paper, we present a simple infinite server queueing model of the number of infected people with Covid-19.  This analysis can be used to understand how many people will become infected, need a hospital bed, or need a ventilator.  We show how the dynamics of the number infected explicitly depends on the arrival rate function.  We explore two arrival rate functions.  The first is symmetric and is given by a Gaussian distribution function and the second is is asymmetric and is given by a gamma distribution function.  We explicitly calculate the time of the peak queue length, the peak queue length, and its dependence on the duration of infection.  Despite the insights derived in this paper, this is not the whole story and many questions still remain.  In what follows, we outline some of the important questions that still remain and need to be studied.  

By leveraging operations research tools and techniques, hospitals have prided themselves on the efficiency that comes with just-in-time supply management and minimizing empty beds, thus saving on hospital costs.  However as a result, they are often not equipped for an epidemic surge such as this one.  Despite our efforts to understand the impact of COVID-19 on the healthcare system, there is plenty of additional research that needs to be done in order to understand emerging complexities.  
One area of importance is to understand the effects of surges in arrivals into the healthcare system.  Some recent research has studied queueing models with self-exciting point processes that model random surges in demand, see for example \citet{daw2018exact, daw2018queue, daw2018queues, pender2016risk, niyirora2016optimal, massey2018dynamic}.  These types of models would generalize our model in this paper, but, they are more complicated and would disguise the easy to understand insights from simpler models.

In this work we have not explicitly analyzed the impact of testing and lags in testing for the coronavirus.  It has been shown through empirical analyses that testing delays can have a huge impact on the reported numbers of positive cases and health outcomes, see for example \citet{Mosley2020}.  We believe that more research in understanding how testing can affect the spread of the disease is important.  It is clear from our cdf plots in Figure \ref{cdfs} that testing is important for predicting deaths that are to follow.  

Finally, we are observing in some cities that it is important to use demographic information for understanding the impact of COVID-19 on particular communities or regions of the country.  In the city of Chicago, it has been observed that even though African Americans make up roughly 30\% of Chicago's population, African Americans make up 52\% of cases of COVID-19 and 68\% of deaths, see for example \citet{black_chicago}.  Also in NYC, it has been reported that COVID-19 is impacting low-income communities much harder than those that are well off \citet{wilsonlowincome}.  Recent work by \citet{kahara2017using} combines queueing theory with demographic information to get a deeper understanding of blood donation dynamics and it is clear that similar demographic analyses are also needed to understand the full impact of COVID-19 on marginalized populations.

\bibliographystyle{plainnat}
\bibliography{exp_growth}

\section{Appendix}

\subsection{The $M_t/G/\infty$ Queue with a Gaussian Arrival Rate}

\subsubsection{Proof of Theorem \ref{time_peak} }
 \begin{proof} 
In order to find the peak queue length we need to set the time derivative of the queue length to zero i.e.
\begin{equation}
\mathop q\limits^ \bullet  (t^*) = 0 .
\end{equation}
Now using the differential equation given in Equation \ref{diffeqn}, we have 
\begin{eqnarray}
\mathop q\limits^ \bullet  (t^*) =  E\left[ {\left( {t^* - \tau  - {S_{\text{e}}}} \right) \cdot \varphi \left( {\frac{{t^* - \tau  - {S_{\text{e}}}}}{\sigma }} \right)} \right] = 0.
\end{eqnarray}
This implies that 
\begin{eqnarray}
\left( {t^* - \tau } \right) \cdot {\text{E}}\left[ {\varphi \left( {\frac{{t^* - \tau  - {S_{\text{e}}}}}{\sigma }} \right)} \right] &=& {\text{E}}\left[ {{S_{\text{e}}} \cdot \varphi \left( {\frac{{t^* - \tau  - {S_{\text{e}}}}}{\sigma }} \right)} \right] .
\end{eqnarray}
Now isolate $t^*$ by itself to get 
\begin{eqnarray}
t^* &=& \tau  + \frac{{{\text{E}}\left[ {{S_{\text{e}}} \cdot \varphi \left( {\frac{{t^* - \tau  - {S_{\text{e}}}}}{\sigma }} \right)} \right]}}{{{\text{E}}\left[ {\varphi \left( {\frac{{t^* - \tau  - {S_{\text{e}}}}}{\sigma }} \right)} \right]}}  \\
&=& \tau  + \frac{{\lambda  \cdot {\text{E}}S \cdot {\text{E}}\left[ {{S_{\text{e}}} \cdot \varphi \left( {\frac{{t^* - \tau  - {S_{\text{e}}}}}{\sigma }} \right)} \right]}}{{\sigma  \cdot \frac{{\lambda  \cdot {\text{E}}S}}{\sigma } \cdot {\text{E}}\left[ {\varphi \left( {\frac{{t^* - \tau  - {S_{\text{e}}}}}{\sigma }} \right)} \right]}} \\
&=& \tau  + \frac{{\lambda  \cdot {\text{E}}S \cdot {\text{E}}\left[ {{S_{\text{e}}} \cdot \varphi \left( {\frac{{t^* - \tau  - {S_{\text{e}}}}}{\sigma }} \right)} \right]}}{{\sigma  \cdot {q_{M/G/\infty }} \left( {t^*} \right)}} .
\end{eqnarray}
This completes the proof.  
\end{proof}

\subsubsection{Discrete Service Distribution}

\begin{corollary}
The $M_t/D_n/\infty$ queueing model with a Gaussian distribution arrival rate is the solution to the following ordinary differential equation 
\begin{eqnarray} \label{diffeqn_2}
 \mathop q\limits^ \bullet  (t) &=&  \sum^{n}_{i=1} p_i \cdot \left( \lambda (t) - \lambda(t-\Delta_i) \right) \\
 &=& \sum^{n}_{i=1} p_i \cdot \left(  \frac{\lambda }{\sigma } \cdot \varphi \left( {\frac{{t - \tau }}{\sigma }} \right) - \frac{\lambda }{\sigma } \cdot \varphi \left( {\frac{{t - \tau - \Delta_i }}{\sigma }} \right)  \right)
 \end{eqnarray}
and the solution is given by
\begin{eqnarray} \label{eqn_soln_2}
{q_{M/G/\infty }}(t) &=& \sum^{n}_{i=1} p_i \cdot \left(   \int^{t}_{(t-\Delta_i)^+} \frac{\lambda }{\sigma } \cdot \varphi \left( {\frac{{s - \tau }}{\sigma }} \right) ds \right) \\
 &=& \sum^{n}_{i=1} p_i \cdot \left( \lambda \left( \mathrm{erf}\left( \frac{ t-\frac{\tau}{\sigma}  }{\sqrt{2}} \right) - \mathrm{erf}\left( \frac{ t - \Delta_i - \frac{\tau}{\sigma} }{\sqrt{2}} \right) \right) \right).
\end{eqnarray}  
%Finally, the time of the peak is equal to 
%\begin{eqnarray} 
%t^* &=& \tau + \frac{\Delta}{2}
%\end{eqnarray}  
%when $ \tau \geq \Delta$.  
\begin{proof}
The proof follows the thinning of Poisson processes and the expression given in Equation \ref{det_soln}.  
\end{proof}
\end{corollary}

\subsubsection{Exponential Service Distribution}

Theorem \ref{exact-lag}
\begin{proof}
First we observe that from Theorem \ref{time_peak} that $t^*$ solves the following fixed point equation
\begin{eqnarray}
t^* &=& \tau  + \frac{{{\text{E}}\left[ {{S_{\text{e}}} \cdot \varphi \left( {\frac{{t^* - \tau  - {S_{\text{e}}}}}{\sigma }} \right)} \right]}}{{{\text{E}}\left[ {\varphi \left( {\frac{{t^* - \tau  - {S_{\text{e}}}}}{\sigma }} \right)} \right]}} .
\end{eqnarray}

Thus, if we define $\mathcal{X}$ to be a unit exponential random variable, then we have 

\begin{eqnarray}
  \ell  \equiv {\text{peak lag}} &=& \frac{{\mathrm{E} \left[ {{S_e} \cdot \varphi \left( {\frac{{\ell  - {S_e}}}{\sigma }} \right)} \right]}}{{\mathrm{E} \left[ {\varphi \left( {\frac{{\ell  - {S_e}}}{\sigma }} \right)} \right]}} 
  \\
  &=& \frac{{\mathrm{E} \left[ {\frac{\mathcal{X}}{\mu } \cdot \varphi \left( {\frac{{\ell  - {\mathcal{X} \mathord{\left/
 {\vphantom {\mathcal{X} \mu }} \right.
 \kern-\nulldelimiterspace} \mu }}}{\sigma }} \right)} \right]}}{{\mathrm{E} \left[ {\varphi \left( {\frac{{\ell  - {\mathcal{X} \mathord{\left/
 {\vphantom {\mathcal{X} \mu }} \right.
 \kern-\nulldelimiterspace} \mu }}}{\sigma }} \right)} \right]}} \\ 
 &=& \frac{{\mu \sigma  \cdot {e^{ - \mu \ell }} \cdot {e^{{\mu ^2}{\ell ^2}/2}} \cdot \left( {\left( {\ell  - \mu {\sigma ^2}} \right) \cdot \Phi \left( {\frac{\ell }{\sigma } - \mu \sigma } \right) + \sigma  \cdot \varphi \left( {\frac{\ell }{\sigma } - \mu \sigma } \right)} \right)}}{{\mu \sigma  \cdot {e^{ - \mu \ell }} \cdot {e^{{\mu ^2}{\ell ^2}/2}} \cdot \Phi \left( {\frac{\ell }{\sigma } - \mu \sigma } \right)}} \\ 
& \Downarrow \nonumber& \\
   \ell  &=& \ell  - \mu {\sigma ^2} + \sigma  \cdot \frac{{\varphi \left( {\frac{\ell }{\sigma } - \mu \sigma } \right)}}{{\Phi \left( {\frac{\ell }{\sigma } - \mu \sigma } \right)}} \\ 
  &  \Downarrow \nonumber & \\ 
     \mu \sigma  &=& \frac{{\varphi \left( {\frac{\ell }{\sigma } - \mu \sigma } \right)}}{{\Phi \left( {\frac{\ell }{\sigma } - \mu \sigma } \right)}} \\
       &  \Downarrow \nonumber & \\ 
        \psi \left( {\mu \sigma } \right) &=& \frac{\ell }{\sigma } - \mu \sigma.  
 \end{eqnarray}
 
 Now it remains to prove the following equalities 
\begin{eqnarray}
  {\mathrm{E} \left[ {\varphi \left( {\frac{{\ell  - {\mathcal{X} \mathord{\left/
 {\vphantom {\mathcal{X} \mu }} \right.
 \kern-\nulldelimiterspace} \mu }}}{\sigma }} \right)} \right]} &=&{\mu \sigma  \cdot {e^{ - \mu \ell }} \cdot {e^{{\mu ^2}{\ell ^2}/2}} \cdot \Phi \left( {\frac{\ell }{\sigma } - \mu \sigma } \right)} 
\\
{\mathrm{E} \left[ {\frac{\mathcal{X}}{\mu } \cdot \varphi \left( {\frac{{\ell  - {\mathcal{X} \mathord{\left/
 {\vphantom {\mathcal{X} \mu }} \right.
 \kern-\nulldelimiterspace} \mu }}}{\sigma }} \right)} \right]}&=&  {\mu \sigma  \cdot {e^{ - \mu \ell }} \cdot {e^{{\mu ^2}{\ell ^2}/2}} \cdot \left( {\left( {\ell  - \mu {\sigma ^2}} \right) \cdot \Phi \left( {\frac{\ell }{\sigma } - \mu \sigma } \right) + \sigma  \cdot \varphi \left( {\frac{\ell }{\sigma } - \mu \sigma } \right)} \right)} .
 \end{eqnarray}

For the first equality we define $\mathcal{G}$ to be a Gaussian random variable and we will exploit the version of Stein's lemma given in \citet{massey2013gaussian, massey2018dynamic} to obtain
\begin{eqnarray*}
  \mathrm{E} \left[ {\varphi \left( {\frac{{\ell  - {\mathcal{X} \mathord{\left/
 {\vphantom {\mathcal{X} \mu }} \right.
 \kern-\nulldelimiterspace} \mu }}}{\sigma }} \right)} \right] &=& \mathrm{E} \left[ {\mathrm{E} \left[ {\left. {\mathcal{G} \cdot \left\{ {\mathcal{G} > \frac{{\ell  - {\mathcal{X} \mathord{\left/
 {\vphantom {\mathcal{X} \mu }} \right.
 \kern-\nulldelimiterspace} \mu }}}{\sigma }} \right\}} \right|\mathcal{X}} \right]} \right] \\
 &=& \mathrm{E} \left[ {\mathrm{E} \left[ {\left. {\mathcal{G} \cdot \left\{ {\mathcal{G} > \frac{{\ell  - {\mathcal{X} \mathord{\left/
 {\vphantom {\mathcal{X} \mu }} \right.
 \kern-\nulldelimiterspace} \mu }}}{\sigma }} \right\}} \right|\mathcal{G}} \right]} \right] \\ 
   &=& \mathrm{E} \left[ {\mathcal{G} \cdot {\text{P}}\left\{ {\left. {\mathcal{G} > \frac{{\ell  - {\mathcal{X} \mathord{\left/
 {\vphantom {\mathcal{X} \mu }} \right.
 \kern-\nulldelimiterspace} \mu }}}{\sigma }} \right|\mathcal{G}} \right\}} \right] \\
 &=& \mathrm{E} \left[ {\mathcal{G} \cdot {\text{P}}\left\{ {\left. {\mathcal{X} > \mu  \cdot \left( {\ell  - \sigma \mathcal{G}} \right)} \right|\mathcal{G}} \right\}} \right] \\ 
   &=& \mathrm{E} \left[ {\mathcal{G} \cdot \left( {{e^{ - \mu  \cdot \left( {\ell  - \sigma \mathcal{G}} \right)}} \cdot \left\{ {\mathcal{G} \leq \frac{\ell }{\sigma }} \right\} + \overline {\left\{ {\mathcal{G} \leq \frac{\ell }{\sigma }} \right\}} } \right)} \right] \\
   &=& \operatorname{E} \left[ {\mathcal{G} \cdot \left( {{e^{ - \mu  \cdot \left( {\ell  - \sigma \mathcal{G}} \right)}} - 1} \right) \cdot \left\{ {\mathcal{G} \leq \frac{\ell }{\sigma }} \right\}} \right] \\ 
   &=& \operatorname{E} \left[ {\mu \sigma  \cdot {e^{ - \mu  \cdot \left( {\ell  - \sigma \mathcal{G}} \right)}} \cdot \left\{ {\mathcal{G} \leq \frac{\ell }{\sigma }} \right\}} \right] \\
   &=& \mu \sigma  \cdot {e^{ - \mu \ell }} \cdot \mathrm{E} \left[ {{e^{\mu \sigma \mathcal{G}}} \cdot \left\{ {\mathcal{G} \leq \frac{\ell }{\sigma }} \right\}} \right] \\ 
   &=& \mu \sigma  \cdot {e^{ - \mu \ell }} \cdot \mathrm{E} \left[ {{e^{\mu \sigma \mathcal{G}}}} \right] \cdot {\text{P}}\left\{ {\mathcal{G} + \mu \sigma  \leq \frac{\ell }{\sigma }} \right\}  \\ 
   &=& {\mu \sigma  \cdot {e^{ - \mu \ell }} \cdot {e^{{\mu ^2}{\sigma ^2}/2}} \cdot \Phi \left( {\frac{\ell }{\sigma } - \mu \sigma } \right)}. 
\end{eqnarray*} 
For the second equality, we have that
\begin{eqnarray*}
  \mathrm{E} \left[ {\frac{\mathcal{X}}{\mu } \cdot \varphi \left( {\frac{{\ell  - {\mathcal{X} \mathord{\left/
 {\vphantom {\mathcal{X} \mu }} \right.
 \kern-\nulldelimiterspace} \mu }}}{\sigma }} \right)} \right] &=& \mathrm{E} \left[ {\mathrm{E} \left[ {\left. {\frac{{\mathcal{X}\mathcal{G}}}{\mu } \cdot \left\{ {\mathcal{G} > \frac{{\ell  - {\mathcal{X} \mathord{\left/
 {\vphantom {\mathcal{X} \mu }} \right.
 \kern-\nulldelimiterspace} \mu }}}{\sigma }} \right\}} \right|\mathcal{X}} \right]} \right] \\
 &=& \mathrm{E} \left[ {\mathrm{E} \left[ {\left. {\frac{{\mathcal{X}\mathcal{G}}}{\mu } \cdot \left\{ {\mathcal{G} > \frac{{\ell  - {\mathcal{X} \mathord{\left/
 {\vphantom {\mathcal{X} \mu }} \right.
 \kern-\nulldelimiterspace} \mu }}}{\sigma }} \right\}} \right|\mathcal{G}} \right]} \right] \\ 
   &=& \mathrm{E} \left[ {\mathcal{G} \cdot \mathrm{E} \left[ {\left. {\frac{\mathcal{X}}{\mu } \cdot \left\{ {\mathcal{G} > \frac{{\ell  - {\mathcal{X} \mathord{\left/
 {\vphantom {\mathcal{X} \mu }} \right.
 \kern-\nulldelimiterspace} \mu }}}{\sigma }} \right\}} \right|\mathcal{G}} \right]} \right] \\
 &=& \mathrm{E} \left[ {\mathcal{G} \cdot \mathrm{E} \left[ {\left. {\frac{\mathcal{X}}{\mu } \cdot \left\{ {\frac{\mathcal{X}}{\mu } > \ell  - \sigma \mathcal{G}} \right\}} \right|\mathcal{G}} \right]} \right] \\ 
   &=& \mathrm{E} \left[ {\mathcal{G} \cdot \left( {\mathrm{E} \left[ {\left. {\frac{\mathcal{X}}{\mu } \cdot \left\{ {\frac{\mathcal{X}}{\mu } > \ell  - \sigma \mathcal{G}} \right\}} \right|\mathcal{G}} \right] \cdot \left\{ {\mathcal{G} \leq \frac{\ell }{\sigma }} \right\} + \mathrm{E} \left[ {\left. {\frac{\mathcal{X}}{\mu }} \right|\mathcal{G}} \right] \cdot \overline {\left\{ {\mathcal{G} \leq \frac{\ell }{\sigma }} \right\}} } \right)} \right] \\ 
   &=& \mathrm{E} \left[ {\mathcal{G} \cdot \left( {{e^{ - \mu  \cdot \left( {\ell  - \sigma \mathcal{G}} \right)}} \cdot \left( {\frac{1}{\mu } + \ell  - \sigma \mathcal{G}} \right) \cdot \left\{ {\mathcal{G} \leq \frac{\ell }{\sigma }} \right\} + \frac{1}{\mu } \cdot \overline {\left\{ {\mathcal{G} \leq \frac{\ell }{\sigma }} \right\}} } \right)} \right] \\ 
   &=& \mathrm{E} \left[ {\mathcal{G} \cdot \left( {{e^{ - \mu  \cdot \left( {\ell  - \sigma \mathcal{G}} \right)}} \cdot \left( {\frac{1}{\mu } + \ell  - \sigma \mathcal{G}} \right) - \frac{1}{\mu }} \right) \cdot \left\{ {\mathcal{G} \leq \frac{\ell }{\sigma }} \right\}} \right] \\ 
   &=& \mathrm{E} \left[ {\left( {{e^{ - \mu  \cdot \left( {\ell  - \sigma \mathcal{G}} \right)}} \cdot \left( {\mu \sigma  \cdot \left( {\frac{1}{\mu } + \ell  - \sigma \mathcal{G}} \right) - \sigma } \right)} \right) \cdot \left\{ {\mathcal{G} \leq \frac{\ell }{\sigma }} \right\}} \right] \\ 
   &=& \mu \sigma  \cdot {e^{ - \mu \ell }} \cdot \mathrm{E} \left[ {{e^{\mu \sigma \mathcal{G}}} \cdot \left( {\ell  - \sigma \mathcal{G}} \right) \cdot \left\{ {\mathcal{G} \leq \frac{\ell }{\sigma }} \right\}} \right] \\ 
   &=& \mu \sigma  \cdot {e^{ - \mu \ell }} \cdot \mathrm{E} \left[ {{e^{\mu \sigma \mathcal{G}}}} \right] \cdot \mathrm{E} \left[ {\left( {\ell  - \sigma  \cdot \left( {\mathcal{G} + \mu \sigma } \right)} \right) \cdot \left\{ {\mathcal{G} + \mu \sigma  \leq \frac{\ell }{\sigma }} \right\}} \right] \\ 
   &=& \mu \sigma  \cdot {e^{ - \mu \ell }} \cdot {e^{{\mu ^2}{\sigma ^2}/2}} \cdot \left( {\left( {\ell  - \mu {\sigma ^2}} \right) \cdot \Phi \left( {\frac{\ell }{\sigma } - \mu \sigma } \right) + \sigma  \cdot \varphi \left( {\frac{\ell }{\sigma } - \mu \sigma } \right)} \right) .
\end{eqnarray*} 
This completes the proof.

\end{proof}

\begin{theorem} \label{peak_bounds}
For the function $\psi(x)$, we have the following bounds.
\begin{eqnarray}
\frac{1}{x} - \frac{1}{x^3 + x} \leq \psi(x) + x \leq \frac{1}{x}.
\end{eqnarray}
Thus, we have in the exponential service setting with a Gaussian arrival rate that the time lag is bounded by 
\begin{eqnarray}
\frac{\sigma}{ \mu \sigma} - \frac{\sigma}{\mu^3 \sigma^3 + \mu \sigma} &\leq& \sigma( \psi( \mu \sigma ) + \mu \sigma ) \leq \frac{\sigma}{\mu \sigma} \\
&& \Downarrow \nonumber 
\end{eqnarray}

\begin{eqnarray}
\frac{1}{ \mu} - \frac{1}{\mu^3 \sigma^2 + \mu } \leq \ell \leq \frac{1}{\mu} = E[S].
\end{eqnarray}

\begin{proof}
The first inequality follows from Theorem 7.2 in \citet{hampshire2009dynamic}.  Therefore, the lag bounds follow by substituting $x = \sigma \mu$.  
\end{proof}

\end{theorem}

\subsubsection{Hyper-Exponential Service Distribution}

Another continuous distribution of interest is the hyper-exponential distribution. The hyper-exponential distribution is a special case of a phase type distribution.  Unlike the Erlang distribution it allows for more variability and has more variance than the exponential distribution.  Usually the hyper-exponential distribution is determined by two M-dimensional vectors of parameters $(p_1, p_2, ... , p_n)$ and $(\mu_1, \mu_2, ..., \mu_n)$. The vector $(p_1, p_2, ... , p_n)$ represents the probabilistic weights of each exponential distribution and the vector  $(\mu_1, \mu_2, ..., \mu_n)$ is the associated vector of rate parameters for each exponential distribution.  Thus, the hyper-exponential is a convex combination of exponential distributions with different rate parameters.  

\begin{corollary}
The $M_t/H_n/\infty$ queue with arrival rate given by a Gaussian distribution with parameters has the following closed form expression for the mean transient queue length 

\begin{eqnarray}
q^{H_n}_{\infty}(t) &=& \sum^{n}_{i=1} p_i \cdot q^{\mu_i}_{\infty}(t)
\end{eqnarray}
where each $q^{\mu_i}_{\infty}(t)$ is equal to
\begin{eqnarray}
q^{\mu_i}_{\infty}(t) &=& \sigma  \cdot e^{ - \mu_i (t - \tau) } \cdot {e^{{\mu_i^2}{\sigma ^2}/2}} \cdot \Phi \left( {\frac{(t - \tau) }{\sigma } - \mu_i \sigma } \right) 
\end{eqnarray}
%\begin{eqnarray}
%q^{\mu_i}_{\infty}(t) &=& q_0 e^{- \mu_i t } +  \frac{ \lambda e^{-\mu_i t } \cdot e^{\mu_i \tau+ \sigma^2 \mu_i^2/2} }{2 \sigma }  \cdot \left( \mathrm{erf}\left( \frac{ t-\frac{\tau}{\sigma} -\sigma \mu_i }{\sqrt{2}} \right) - \mathrm{erf}\left( \frac{ -\frac{\tau}{\sigma} -\sigma \mu_i}{\sqrt{2}} \right) \right).
%\end{eqnarray}

\begin{proof}
This follows from the fact that the hyper-exponential is just a thinning of a Poisson process.  Thus, the sum is also a Poisson process with the sum of the rates.  
\end{proof}

\end{corollary}

\begin{corollary}
The $M_t/H_n/\infty$ queue with arrival rate given by a Gaussian distribution with parameters has the following closed form expression for the lag

\begin{eqnarray}
\ell  &=&\frac{\sum^{n}_{j=1} \alpha_j {\mu_j \sigma  \cdot {e^{ - \mu_j \ell }} \cdot {e^{{\mu_j^2}{\sigma ^2}/2}} \cdot \Phi \left( {\frac{\ell }{\sigma } - \mu_j \sigma } \right)}  }{\sum^{n}_{j=1}   \alpha_j  \mu_j \sigma  \cdot {e^{ - \mu_j \ell }} \cdot {e^{{\mu_j^2}{\sigma ^2}/2}} \cdot \left( {\left( {\ell  - \mu_j {\sigma ^2}} \right) \cdot \Phi \left( {\frac{\ell }{\sigma } - \mu_j \sigma } \right) + \sigma  \cdot \varphi \left( {\frac{\ell }{\sigma } - \mu_j \sigma } \right)} \right)}
\end{eqnarray}
where 
\begin{eqnarray}
\alpha_j &=& \frac{ \left( \prod^{n}_{i=1} \mu_i \right) p_j }{ \sum^{n}_{i=1} p_i \left( \prod^{n}_{i \neq j} \mu_i \right) } 
\end{eqnarray}

\begin{proof}
This follows from the fact that the hyper-exponential is just a thinning of a Poisson process.  Thus, the sum is also a Poisson process with the sum of the rates.  Moreover, the stationary excess distribution of a hyper-exponential is another hyper-exponential distribution with modified rates.  
\end{proof}

\end{corollary}

\end{document}